\documentclass[manuscript]{aastex}

\newcommand{\be}{\begin{equation}}
\newcommand{\ee}{\end{equation}}
\newcommand{\bea}{\begin{eqnarray}}
\newcommand{\eea}{\end{eqnarray}}
\newcommand{\bean}{\begin{eqnarray*}}
\newcommand{\eean}{\end{eqnarray*}}

\begin{document}
\title{MHD WAVES AND CORONAL HEATING: UNIFYING EMPIRICAL AND MHD
  TURBULENCE MODELS}

\author{Igor V. Sokolov\altaffilmark{1}, Bart van der
  Holst\altaffilmark{1}, Rona Oran\altaffilmark{1}, Cooper Downs\altaffilmark{2}, Ilia I. Roussev\altaffilmark{2},
Meng Jin\altaffilmark{1}, Ward B. Manchester IV\altaffilmark{1},
Rebekah M. Evans\altaffilmark{3},
and Tamas I. Gombosi\altaffilmark{1}}

\altaffiltext{1}{Department of AOSS, University of Michigan,
2455 Hayward St, Ann Arbor, MI 48109; igorsok@umich.edu.}
\altaffiltext{2}{Institute for Astronomy, University of Hawaii,
2680 Woodlawn Dr, Honolulu, HI 96822.}
\altaffiltext{3}{Goddard Space Flight Center, Greenbelt, MD.}

\begin{abstract}
We present a new global model of the solar corona,
including the low corona, the transition region and the top of
chromosphere. The realistic 3D magnetic field is simulated using the
data from the photospheric magnetic field measurements. The distinctive
feature of the new model is incorporating the MHD Alfven wave
turbulence. We assume this turbulence and its non-linear dissipation 
to be the only momentum and energy source for heating the coronal
plasma and driving the solar wind. The difference between the turbulence
dissipation efficiency in coronal holes and that in closed field regions
is because the non-linear cascade rate degrades in strongly
anisotropic (imbalanced) turbulence in coronal holes (no inward propagating
wave), thus resulting in colder coronal holes with the bi-modal
solar wind originating from them. The detailed presentation of the
theoretical model is illustrated with the synthetic images for
multi-wavelength EUV emission compared with the observations from SDO
AIA and Stereo EUVI instruments for the Carrington rotation 2107.

\end{abstract}



%

\section{Introduction}
\label{Intro}
Results from Hinode observations have recently upped the estimates for the MHD wave
energetics in the solar corona \citep{dupo08}. Observed in the chromosphere, the
magnetic field perturbations appear to be so powerful that even a
$10\div20\%$ fraction of them, which propagate out from the Sun, carry a
large enough energy to heat the solar corona and accelerate the solar wind.
Even more promising, with the launch of the Solar Dynamics 
Observatory, we are now beginning to see observational hints of these ubiquitous 
waves in the transition region and low corona \citep{mcintosh11}.

Even before these encouraging observations had been
obtained, models which incorporated, or even were entirely based
upon Alfv\'en wave turbulence as the momentum and energy
source were developed to describe the solar wind and
coronal heating. Nowadays, {\it developing
  the turbulence-driven global space weather model} becomes a problem
tempting to be solved.     
\subsection{Solar wind: can the turbulence-driven model compete with
  the semi-empirical one?}
In \cite{usma00}, a three-dimensional (3D)
model for the solar wind was suggested in which the Alfv\'en
wave turbulence pressure served to accelerate the solar wind. 
The solar wind bi-modal structure as observed by Ulysses was
successfully reproduced in the numerical simulation. However, the
quantitative agreement of this model with
long history of the solar wind observations at 1 AU is insufficient for 
global space weather simulations.  
The same criticism seems to be applicable to the more refined
and physics-based  Alfv\'en-wave-driven models of the solar wind 
\cite{suzu05,verd10,osman11}. 

Therefore, the semi-empirical approach so far is better suited for
global space weather simulations. The most popular parameterization was
adopted by \cite{arge00} in their Wang-Sheeley-Arge (WSA) model,
which well describes the solar wind parameters at 1~AU.  The
semi-empirical ``synoptic'' formulae for the solar wind speed employ
the solar magnetogram and the properties of the magnetic lines of
the {\em potential magnetic field} as recovered from the synoptic
magnetogram data.  The empirical dependence of the solar wind
properties from two input parameters: $\theta$, which is the angular
distance from the solar wind origin point till the nearest coronal hole
boundary, and $f_{\rm exp}$, which is the expansion factor, $f_{\rm exp}=|{\bf B}_{R =
R_\odot}|R_s^2/[|{\bf B}_{R=2.5R_\odot}|(2.5R_s)^2]$, is derived from
the following two assumptions and  observations.  First, the slow
solar wind is assumed to originate from the coronal hole boundary
(small values of $\theta$), whereas the fast solar wind originates
from the central part of the coronal hole (large values of $\theta$).
Second, small coronal holes (having large values of $f_{\rm exp}$) usually
produce slower solar wind compared to large coronal holes (having
small values of $f_{\rm exp}$). 

In \cite{cohen07}, we coupled the semi-empirical WSA
formulae to the global three-dimensional model for the solar corona
and inner heliosphere within the Space Weather Modeling Framework
(SWMF) \cite[]{toth05}.  The WSA formulae were used as the boundary
condition  for the model which had been coupled to the MHD simulator
via the varied polytropic index distribution (see \cite{roussev03b}).
However, the physics of the Alfv\'en wave turbulence has almost no
intersection with the semi-empirical model of WSA. 

{\it In order to be competitive with semi-empirical
  model of the solar wind, the turbulence-driven model should
  quantitatively reproduce the solar wind variation from the coronal
  hole boundary to the coronal hole central part, similar to that
  parameterized by the expansion factor.}
\subsection{Coronal heating: can the turbulence-based model compete
  with {\it ad hoc} heating functions?}
Proceeding from the solar wind to the coronal heating, we can see again
the disconnection between the physics-based models for the turbulent
heating in the corona, one of the most advanced models of this kind
being recently described in \cite{cran10} (see also \cite{tu97,hu00,Habb03,dmit02}) on one hand, and well 
established models with semi-empirical heating function, such as that
presented in \cite{lion01,rile06,tito08,lion09,downs10}, on the other
hand. The heating function is applied to power the plasma in the solar corona
with some heating rate, the distribution of which is given by some
functions chosen not from some deep physical considerations, but
chosen in an ad-hoc manner to provide a better agreement with
observations.  

These heating functions should be applied in conjunction with
thermodynamic energy equation(s) in the 
Lower Corona (LC) model.  A direct accounting is needed of the transition region
between the chromosphere and the corona, where non-ideal-MHD thermodynamic 
terms of energy transport, such as electron heat conduction, radiative 
losses, and coronal heating all become important. In \citet{downs10} we
used this thermodynamic MHD model to explore 
empirical parameterizations of coronal heating in the context of 
realistic 3D magnetic structures observed in the EUV and soft-X-Rays 
on Aug 27, 1996. Through direct comparison of synthetic observables 
to observations we demonstrated that 
this model can effectively capture the interplay between coronal
heating and electron heat conduction. 
 
The plasma parameters distribution obtained in this way
may be successfully applied to generate synthetic EUV and X-ray
images, which appear to be in a good agreement with those obtained
with the EIT telescope onboard SOHO and SXT onboard Yohkoh (up to 2001). The
observation synthesis capability has been extended to the major low
corona imaging instruments available in space, namely STEREO/EUVI,
SDO/AIA, and Hinode/XRT. The best agreement with observations is
achieved with {\it ad hoc} heating functions (such as the
unsigned-flux-based heating model as discussed in Section 2.1).  {\it In order to be competitive with ad hoc
  coronal heating models, the turbulence-driven model should
  quantitatively reproduce some successful heating function, that is
  the realistic wave dissipation should provide the same heating rate,
  as, for example, the unsigned-flux-based heating model \cite{abbe07}.}
\subsection{Towards a Global Alfven-Wave-Turbulence-Driven MHD Model of the Solar Corona and Solar Wind}
\label{Instro_SC}
In addition to the
mentioned disconnection between physics-based and observations-driven
models, there are two more contradictions to comment on. First, the
quantitative Alfv\'en-wave-driven models for the solar wind and for the
coronal heating do not well conform with each other.  For example, the
the coronal heating model of \cite{cran10}, once applied to
realistic 3D global model does not give realistic plasma parameters
in the solar wind region (in the coronal holes).  Second, the increased
estimates for the Alfv\'en wave turbulence energetics in the Sun's
proximity require us to revisit the models for the evolution of turbulence
while the solar wind propagates towards 1~AU.  The revisited model
should account for both {\em the observed level and frequency
spectrum of turbulence at 1~AU and the solar wind ion temperature
resulting from the turbulence dissipation.} 

With the advent of modern computational tools it is now becoming the
norm to employ detailed 3D computer models as 
simulation tools that directly account for the inhomogeneous nature of
the Sun-Heliosphere environment. The key advantage of this approach is 
the ability to compare and validate model results through direct 
comparisons to all kinds of the  observational data listed above: for
the solar wind and turbulence characteristics at 1 AU (see also \cite{jin12}) 
and for EUV and
X-ray images of the solar corona.  
\subsection{Goal and Content of the Paper}
In the current paper, we get rid of the {\it ad hoc} heating functions
and parameterize the coronal heating in the LC in terms of the Alfven
wave turbulence dissipation. 

The theoretical model for this approach is summarized in Section 2. In
subsections 2.1, 2.3 we demonstrate the possibility to parameterize a
popular and successful semi-empirical heating model based on unsigned magnetic
flux in terms of the Alfv\'en wave turbulence dissipation. While choosing
the way to parameterize heating, one may try to vary both the boundary
condition for the wave energy flux (the Poynting flux) inflowing to the
solar corona from the solar surface and
the dissipation length for turbulence. In subsection 2.2 we show that
the boundary condition is strongly restricted which reduces
the model uncertainty. With the pre-specified boundary condition, the
drastic difference in the plasma heating mechanism between open and closed
field regions should be caused by a difference in the
wave dissipation efficiency. In subsection 2.4 we suggest and describe
the physical mechanism of a nonlinear interaction between {\it
  oppositely propagating waves}. The degraded intensity of the inward
propagating waves may be responsible for the
reduction in the turbulence dissipation rate in the coronal holes thus 
resulting in the bimodal solar wind structure. 

In Section 3 we summarize a computational model and the code we use
to simulate the state of the solar corona. Section 4 presents the
simulation results for CR2107 and their comparison with EUV images. In
the Conclusion we discuss the plans for a future.  
\section{Coronal Heating and Its Parameterization
via the Alfv\'en Wave Turbulence}
\subsection{Coronal Heating Model Based on
the Unsigned Magnetic Flux}
Among the possible choices for the volumetric heating function
examined in \cite{downs10}, the most advantageous one 
reproduces both EUV and soft X-ray observations was that adopted
earlier on by \cite{abbe07}.  This heating function is based on the
scaling law obtained by \cite{Pevtsov}.  This law establishes the
power-law relationship between the total heating power, $E =
\int{e dV}$, integrated over the plasma volume above $1 \, R_\odot$,
and the unsigned magnetic flux, $\Phi = \int{|{\bf B}_{R_\odot}
\cdot {\bf dS}|}$, integrated over the photospheric surface, i.e.,
\begin{equation}\label{eq:Pevtsov}
\int_{R \ge R_\odot}{e dV} = {\rm const} \, \Phi, \quad \Phi =
\int_{R=R_\odot}{|{\bf B} \cdot {\bf dS}|}.
\end{equation}
Here $R$ is the heliocentric distance and $R_\odot$ is the solar radius.
In the original work by \cite{Pevtsov}, the relationship was found not
to be linear: $E_{X} \approx 0.894 \Phi^{1.1488}$ [CGSE].  Note, however,
that the observational data used to derive the scaling law were the
X-Ray total luminosity, $E_X$, which is only a small fraction of the total
heating power, $E$.  There are also losses due to the electron heat
conduction, radiation in spectral ranges other than X-rays, and also
solar wind expansion.  As a result, a scaling factor that relates $E_{X}$
to $E$ is required, $\zeta = E_{X} / E \approx (1\div2) \cdot 10^{-2}$.  If
one assumes that $E_{X} / E$ ratio is somewhat elevated during solar
maximum, or near active regions where the average magnetic field
strength, $|B|$, is higher, so that $\zeta \propto \Phi^{0.1488}$, then
we can state that Eq.(\ref{eq:Pevtsov}) is equivalent to that presented
in \cite{abbe07,Pevtsov}.  On one hand, this relationship quantifies
the well-known fact that during solar maximum, or near active regions,
the average $B$ is higher.  On the other hand, the coronal heating
is more intense in this case.

The constraint given by Eq.(\ref{eq:Pevtsov}) is not sufficient to
establish the 3D distribution of the heating function.  In \cite{abbe07},
the heating function was scaled with the magnetic field: $e \propto
|{\bf B}|$, with the constant factor being chosen to satisfy a power-law
scaling similar to Eq.(\ref{eq:Pevtsov}).  In \cite{downs10} an exponential
envelope has been adopted: $e \propto |{\bf B}| \exp[-(R-R_\odot)/L]$,
where the dissipation length was chosen $L\approx 40$~Mm (herewith, $R$ is the 
heliocentric distance).  The
latter formula, however, appears to be applicable only for closed field
regions away from active regions.  In coronal holes, in turn, the
dissipation length was chosen to be by a factor of ten greater.  In all
the cases, the sophisticated envelope function had to be integrated
over the volume in order to obtain the common constant factor
satisfying Eq.(\ref{eq:Pevtsov}).

Below we demonstrate that a model of coronal heating that both satisfies
Eq.(\ref{eq:Pevtsov}) and provides the spatial distribution of the heating
envelope similar to that discussed above can be realized assuming 
Alfv\'en waves being the main heating source the solar corona.  Without
going into details about the wave absorption and non-linear conversion
and transport, we can construct the heating function by:  (i) imposing a
boundary condition for the Poynting flux of Alfv\'en wave energy; and,
(ii) adopting an absorption mechanisms for the waves that would result
in the desired distribution of energy deposition from waves into the
coronal plasma.
\subsection{``Percolation'' Boundary Condition for the
Poynting Flux at the Photosphere}
Assuming that the Alfv\'en wave turbulence is the main source for
heating the coronal plasma, here we discuss a choice of boundary
condition for their Poynting vector, $P$ at the solar surface.  We
consider that the local value of the Pointing flux at the photosphere,
$P_{R_\odot}$, scales with the magnetic field magnitude,
$| {\bf B}|_{R_\odot}$ as:
\begin{equation}\label{eq:crazy}
P_{R_\odot} = {\rm const} \,| {\bf B}|_{R_\odot}.
\end{equation} 
With this choice for $P$, Eq.(\ref{eq:Pevtsov}) is automatically
satisfied if one assumes that the entire energy of the Alfv\'en waves
traveling from the photosphere to the solar corona: 
$\int_{R = R_\odot}{\frac{P_{R_\odot}}{|{\bf B}|_{R_\odot}}|{\bf B}\cdot{\bf dS}|}$,
is deposited into the coronal plasma, $\int_{R\ge R_\odot}{e\,dV}$.
Then:
$$
E=\int_{R \ge R_\odot}{e\,dV} = \int_{R=R_\odot} {\frac{P_{R_\odot}}{|{\bf B}
|_{R_\odot}}|{\bf B}\cdot{\bf dS}|} =
$$
$$
= \frac{P_{R_\odot}}{|{\bf B}|_{R_\odot}} \int_{R=R_\odot}{|{\bf B} \cdot
{\bf dS}|} = \frac{P_{R_\odot}}{|{\bf B}|_{R_\odot}}\Phi,
$$
meaning that Eq.(\ref{eq:Pevtsov}) is fulfilled, and that the exact
same constant is present in Eqs.(\ref{eq:Pevtsov},\ref{eq:crazy}):
\begin{equation}
\frac{E}{\Phi} = \frac{P_{R_\odot}}{|{\bf B} |_{R_\odot}}.
\end{equation} 

However, the relationship given by Eq.(\ref{eq:Pevtsov}) is not the
only way to arrive at Eq.(\ref{eq:crazy}).  It can be also thought of
as the {\em percolation} boundary condition for the Poynting flux
of the waves.  Assume that all the magnetic field in the solar corona
and all the Alfv\'en wave turbulence percolate to the solar corona
from a depth at the solar interior where both the magnetic field
intensity (presumably $50-70$ kG) and the Alfv\'en wave Poynting
flux are very high compared to their photospheric values (except in
sunspots).  In this case, both the magnetic field flux and the wave
energy flux ``propagate'' along the flux tube, and they both vary
inversely proportional to the flux tube cross-section.  (Note that the
growth of the flux tube cross-section from the solar interior towards
the solar surface may be quite large and is hardly predictable.)  As
a consequence, the local value of the magnetic field at the
photosphere is orders of magnitude smaller than that in the solar
interior, and its variation along the solar surface may be quite large.
The same behavior should be true for the Poynting flux.  However,
the {\em ratio} of the Poynting flux to the magnetic field intensity
{\em remains constant along the magnetic tube}, and that is exactly
what is assumed in Eq.(\ref{eq:crazy}).

Surprisingly, in addition to these two considerations, we can cite
four different groups of authors, which arrive at the same condition
as in Eq.(\ref{eq:crazy}) by quite different reasons (and stemming
from quite different observations). First, the assumption as in
Eq.(\ref{eq:crazy}) applied to the radial components of the Poynting
flux, $P_R$, and the magnetic field, $B_R$, is a keystone of the Fisk theory for the
solar wind (see \cite{fisk96,fisk99b,fisk99c,fisk01a,fisk01b}. 
Then, \cite{farr97} (and the works cited
therein) formulated an ``abnormal'' adiabatic expansion law relating the
(turbulent) energy density, $w$ to the mass density, $\rho$: 
$
w\propto \rho^{1/2}$ for the coronal matter, based on numerous
observations relating to the coronal plasma and the CME ejecta. 
The suggested adiabatic law has an abnormal value of the adiabatic
index: $w\propto \rho^\gamma,\,\gamma=\frac12<1$.  Here, assuming
a plasma dominated by turbulence, $\rho$ is the mass density and
$w$ could be the energy density of the Alfv\'en waves ($w \sim (\delta
B)^2$, with $\delta B$ being the irregular magnetic field).  As long as
for the Alfv\'en waves $P=V_Aw$, and the Alfv\'en wave speed scales
with the density and magnetic field as $V_A=\frac{|{\bf B}|}{\sqrt{\mu_0\rho}}
\propto B\rho^{-1/2}$, Eq.(\ref{eq:crazy}) may be also rewritten in the
form of the ``abnormal'' adiabatic law, 
\begin{equation}\label{eq:w}
w_{R_\odot}={\rm const}\sqrt{\mu_0\rho}=\frac{P}{|{\bf B}|}\sqrt{\mu_0\rho}.
\end{equation}
As long as media with adiabatic index less
than one are not thermodynamically stable, it is hardly instructive to
interpret Eq.(\ref{eq:crazy}) in this manner.  However, this unexpected
support for Eq.(\ref{eq:crazy}) is worth mentioning here.

Another example supporting Eq.(\ref{eq:crazy}) is the paper by
\cite{suzu06}, which stems from the Wang-Sheeley-Arge model, and
proposes a semi-empirical quantitative model for the solar wind.  In
their work, it is assumed that the solar wind is mostly powered by the
Alfv\'en wave turbulence, and the input parameter in the model is the
constant value of the following average: $<\delta{\bf B}_\perp\cdot
\delta{\bf v}_\perp>\approx0.83\,{\rm T\,m/s}$ at the solar
surface.  Here, $\delta{\bf B}_\perp$ and $\delta {\bf v}_\perp$ are the
turbulent magnetic field and the turbulent velocity pulsation, respectively.
These are both being orthogonal to the regular magnetic field.  It is
straightforward to demonstrate that the above assumption is
equivalent to Eq.(\ref{eq:crazy}):
$$
 <\delta{\bf
  B}_\perp\cdot\delta{\bf v}_\perp>=
\frac{|{\bf
     B}|}{V_A}\frac{w}\rho=\frac{|{\bf
     B}|^2}{\rho V_A^2}\frac{V_A w}{|{\bf B}|}=
\mu_0\frac{P_{R_\odot}}{|{\bf B}|_{R_\odot}}.
$$
From here, one can see that the Alfv\'en wave turbulence may be
employed to reproduce the observed solar wind parameters with
the use of a boundary condition similar to that given by
Eq.(\ref{eq:crazy}).  In this case, the constraint for the constant
factor in Eqs.(\ref{eq:Pevtsov},\ref{eq:crazy}) is $(P/|{\bf B}|)_{R_\odot}
\ge (0.83\,{\rm T\,m/s})/\mu_0\approx7\cdot 10^5\,
{\rm W/(m^2T)}$.  This is the lower bound as long as in
the model of \cite{suzu06} there is
no energy loss mechanism from the solar wind plasma (e.g. radiation, heat
conduction, etc.). Therefore, the realistic
heating rate should be higher in order to balance these losses.
Nevertheless, the estimate is very close to that which follows from
\cite{Pevtsov,abbe07,downs10}.  Specifically, for the choice $\zeta
=2\cdot10^{-2}\left<|{\bf B}|{\rm [G]}\right>^{0.1488}$, where the average
magnetic field intensity over the solar surface is introduced, the
estimate of the Pointing-flux-to-field ratio is as follows:
$$
\frac{E}{\Phi}=\frac{E_X}{\Phi\zeta}\left(\approx\frac{0.894 \Phi^{1.1488}}{2\cdot10^{-2}|{\bf B}|^{0.1488}\Phi}\approx44.7(4\pi
R_\odot^2)^{0.1488}{\rm [CGSE]}\right)\approx
1.1\cdot10^6\frac{\rm W}{\rm m^2T},
$$  
or, in all equivalent forms:
$$
P\approx1.1\cdot10^6\frac{\rm W}{\rm m^2}\left[\frac{|B|}{1\,T}\right],\quad 
P_R\approx1.1\cdot10^6\frac{\rm W}{\rm m^2}\left[\frac{|B_R|}{1\,T}\right],
$$ 
\begin{equation}\label{eq:bc}
w=\frac{P}{V_A}\approx6.6\cdot10^{-3}{\rm W/m^3}\sqrt{\frac{\rho}{3\cdot 10^{-11}{\rm
      kg/m}^3}},\quad<\delta{\bf
  B}_\perp\cdot\delta{\bf v}_\perp>\approx1.4{\rm Tm/s}
\end{equation}
$$
<\delta{\bf v}_\perp\cdot\delta{\bf v}_\perp>\approx(15\,{\rm
  km/s})^2\sqrt{\frac{3\cdot 10^{-11}{\rm kg/m}^3}{\rho}},
$$ 
where we chose the value of mass density $3\cdot10^{-11}{\rm kg/m^3}$ 
($N_a\approx2\cdot10^{16}{\rm m^{-3}}$) corresponding to the top of
the chromosphere (near the transition region), to compare with the 
Hinode observations of the turbulent velocities $\approx 15$ km/s.  
Thus, a variety of models, which are validated against EUV
observations, X-ray radiance measurements, as well as in-situ solar
wind observations are {\em all in agreement} with Eq.(\ref{eq:crazy}),
with a rather narrow range for the constant factor in the expression.

Next, let us compare this boundary condition with a coronal heating
model based on Alfv\'en wave turbulence, such as the advanced model
of \cite{cran10} (see also the works cited therein).  By comparing
Eq.(\ref{eq:crazy}) with Eq.(29) from \cite{cran10}, one can see that
Eq.(\ref{eq:crazy}) is adopted in their model not just as a boundary
condition at the solar surface, but as a fundamental relationship valid
throughout the solar corona (or, at least, for heliocentric distances
$1.1R_\odot\le R\le 1.5 R_\odot$).  The observational constraints for the
constant factor in Eq.(\ref{eq:crazy}) are somewhat weaker than
the values discussed here ($P/|{\bf B}|\sim(0.5\div0.8\div1.0)\cdot 10^6
{\rm W/(m^2T)}$).  However, the deviation is very small.  In
addition, there is some uncertainty in these observations (e.g.,
the Pointing flux is not measured, but rather the oscillating velocity).
Also, the solar observations at $1.1R_\odot$ cannot be easily converted
into a boundary condition at the photosphere.  Furthermore, the
dissipation length of 40~Mm assumed by \cite{downs10} is shorter
than the difference between 1.1 $R_\odot$ (where the oscillations are
observed) and 1 $R_\odot$ (where the boundary condition is imposed).
As a consequence, it is likely that the turbulence observed at 1.1 $R_\odot$ is already
weakened by the wave absorption.
\subsection{Heating Function Parameterized in Terms of Alfven
  Wave Dissipation}
In order to proceed to the 3D distribution of the heating function, we
represent the volume integral of the heating function as the total of
integrals over the flux tubes:
$$
E=\int{\left(\int{edS(\ell)}\right)d\ell},
$$ 
with $d\ell$ being the length differential along the flux tube, and $dS(\ell)$
being the expanding cross-section of the flux tube.  Along the flux
tube, the magnetic flux is conserved, and hence: $dS(\ell)|{\bf B}|(\ell)=
{\rm const}=|({\bf B}\cdot{\bf dS})|$.  Again, assuming that the Alfv\'en
waves propagate only along the flux tube, their energy conservation
is given by the following relationship: $P(\ell+d\ell)dS(\ell+d\ell)-P(\ell)dS(\ell)=-
ed\ell dS(\ell)$, hence:
\begin{equation}\label{eq:along}
\frac{e(\ell)}
{|{\bf B}|(\ell)}=-\frac\partial{\partial
    \ell}\left(\frac{P(\ell)}{|{\bf B}|(\ell)}\right).
\end{equation}   
By introducing the absorption length, $L$, such that $e=V_Aw/L=P/L$,
one can integrate Eq.(\ref{eq:along}) along the magnetic field line:
$$
P(\ell)=\left(\frac{P}{|{\bf B}|}\right)_{R=R_\odot}|{\bf
B}|(\ell)\exp\left(-\int_0^\ell{\frac{d\ell}{L}}\right),
$$
\begin{equation}
e(\ell)=\left(\frac{P}{|{\bf B}|}\right)_{R=R_\odot}\frac{|{\bf
B}|(\ell)}{L}\exp\left(-\int_0^\ell{\frac{d\ell}{L}}\right).
\end{equation}
Depending on the choice of the absorption length, one can obtain the
following spatial distribution of the heating function: (i) $l\ll L$,
the heating function is the same as that as used in \cite{abbe07}, i.e.,
$e \propto |{\bf B}|$; (ii) for $l\gg L$, the heating function decays
exponentially, i.e., $e\propto\exp(-\int{d\ell/L})$.  We now arrive at an
important conclusion, that is, {\em the most common heating functions
may be parameterized in terms of an effective absorption coefficients
for the Alfv\'en wave turbulence!}  The desired absorption coefficient,
$V_A/L$, may be directly included to the WKB equation for total wave
intensities, $w_{\pm}$, propagating parallel and antiparallel to the
mean magnetic field vector, ${\bf b}={\bf B}/|{\bf B}|$, correspondingly
\citep{usma00}:
\begin{equation}
\label{eq:usma}
\frac{\partial w_{\pm}}{\partial t}+\nabla\cdot\left({\bf
  u}w_{\pm}\pm{\bf b}V_Aw_{\pm}\right)+\frac12w_{\pm}(\nabla\cdot{\bf
  u})=-\frac{V_Aw_{\pm}}L.
\end{equation}
In the open field region (in the solar wind) only outward propagation waves 
were involved in the \citep{usma00} heliosphere model, that is the ``plus'' 
waves in the regions of the solar wind with the positive radial magnetic 
field and the ``minus'' waves elsewhere. In addition to this, the low solar 
corona model of the present paper also includes the 
closed field regions, where the waves of both directions are present emitted  
from the opposite footpoints of a closed magnetic field line. 
However, Eq.(\ref{eq:usma}) can only be applied as a rather crude 
approximation. In the more refined model as derived in \cite{sokolov09})
on the left hand side of Eq.(\ref{eq:usma}) the term proportional to
the velocity divergence and spectral evolution in the
convergent/divergent flows should be properly accounted for. Even more 
important is the role played by the frequency dependence on the right hand
side of Eq.(\ref{eq:usma}):
\begin{equation}\label{soko09}
\frac{\partial I_{\pm}}{\partial t}+\nabla\cdot\left({\bf
  u}I_{\pm}\pm{\bf b}V_AI_{\pm}\right)-\frac\omega2\frac{\partial
  I_{\pm}}{\partial \omega}(\nabla\cdot{\bf
  u})=-\Gamma(\omega)I_\pm,
\end{equation}
where  the equations are formulated for the wave energy
density, $I_\pm$, related to the unit volume and to the interval of the wave
circular frequency, $d\omega$, in the co-moving frame of
reference, so that $w_\pm=\int_0^\infty{I_\pm d\omega}$. 
Not only is the dependence of the dissipation coefficient,
$\Gamma(\omega)$, 
 on the wave frequency significant (practically so that only
for the highest wave frequencies is the absorption non-negligible), but 
also the nonlinear wave-wave interaction may occur in the form of the
wave package upshift in frequency. As a result, the wave energy from
large-scale perturbations flows through the spectrum towards the
short-scale end and once transferred to the shortest possible scales,
the energy dissipates.
\subsubsection{Kolmogorov Turbulence and Dissipation}
We found above that the unsigned-flux-based heating function with the
exponential envelope function can be realized as the dissipation in the
Alfven wave turbulence. In \citep{downs10} we saw that the
unsigned-flux-based heating function well reproduces EUV and X-ray
images of the closed field regions (about the coronal holes, see below)
with the choice for the dissipation length to be $\approx40$ Mm. Here
we discuss a choice for the dissipation length, $L$, based on the
phenomenological turbulence theory, desiring that in the closed field
region $L$ would be about $40$ Mm and do not vary too strongly, at
least at low altitudes $\le 0.1 R_{\odot}\approx 70$ Mm.

Consider extra terms in the governing equations for the Alfven,
turbulent energy density, accounting for the wave-wave 'cascade'
rate. Only the cascade-describing terms are kept. For the closed field
region we employ the key assumption about the {\it isotropic (balanced)} 
turbulence:
\begin{equation}\label{eq:isotr}
I_+=I_-=I.
\end{equation}
For simplicity, assume that
$\Gamma_{ww}(\omega)$ characterizes the rate of conversion of two wave
``photons'' of the frequency of $\sim\omega$ to a single photon with
the frequency, $\sim2\omega$. As long as the wave energy conserves in
this processes, this
may be thought of as the conservative advection in the frequency
space. The positive speed of this advection being approximately
$\sim\omega\Gamma_{ww}(\omega)$, as long as with the rate, $\Gamma_{ww}$ the wave
frequency is upshifted by $[(\sim2\omega)-(\sim\omega)]$. So, we have a
phenomenological equation as follows:
\begin{equation}\label{gen}
\frac{\partial I}{\partial
  t}+
\frac{\partial}{\partial\omega}
\left(\omega\Gamma_{ww}(\omega)I\right)=-\Gamma(\omega) I
\end{equation} 
The dimensionless ratio, $\Gamma_{ww}/\omega$, may be parameterized in
terms of another dimensionless ratio, $\omega I\mu_0/B^2$, $B$ being the
magnetic field magnitude. For two most natural
estimates, $\Gamma_{ww}\propto\omega\cdot(\omega I\mu_0/B^2)\propto\omega (\delta
B/B)^2$ or $\Gamma_{ww}\propto\omega\sqrt{\omega I\mu_0/B^2}\propto\omega (\delta
B/B)$, we arrive at the Kraichnan spectrum or the Kolmogorov spectrum (see \cite{li11}) as
the steady-state solution of Eq.(\ref{gen}) throughout the inertial
range of frequencies (we neglect wave absorption in this range):
$$
\frac{\partial}{\partial\omega}\left[\omega^2\left(\frac{\omega I\mu_0}{B^2}\right)^{0.5;1}I\right]=0
$$ for $I\propto\omega^{-\frac53;-\frac32}$. We chose the
assumption of the Kolmogorov spectrum, so that 
$$\Gamma_{ww}\propto\frac\omega{V_A}\sqrt{\frac{\omega I}{\rho}}.$$
On integrating Eq.(\ref{gen}) over frequency from some frequency value
within the inertial range, $\omega_{in}$, to infinity with neglecting 
wave dissipation within the inertial range, we find:
\begin{equation}
\int_{\omega_{in}}^\infty{\Gamma(\omega) I d\omega}\propto \left(\frac{\omega^2I}{V_A}\sqrt{\frac{\omega
    I}{\rho}}\right)_{\omega=\omega_{in}}.
\end{equation}
The equation is valid for any choice of $\omega_{in}$ within the
inertial range, as long as the right hand side is constant for
$I\propto\omega^{-5/3}$ and $\Gamma(\omega_{in})=0$. The energy of the 
Kolmogorov spectrum, $\propto\int{\omega^{-5/3}d\omega}$, is
  concentrated at low frequency, therefore, at the least possible
  $\omega_{in}\sim\omega_{min}$ one can estimate 
$(\omega I)_{\omega=\omega_{min}}\sim w$ and 
$$
\int_{\omega_{in}}^\infty{\Gamma(\omega) I d\omega}\propto \frac{\omega_{min}}{V_A}\sqrt{\frac{w}\rho}w,
$$  
Following \cite{hollw86}, for an isotropic Alfven wave turbulence we accept 
\begin{equation}
\int_{\omega_{in}}^\infty{\Gamma(\omega) I
  d\omega}\approx\frac1{L_\perp}\sqrt{\frac{w}\rho}w=\frac{\sqrt{<\delta
    {\bf v}_\perp\cdot\delta{\bf v}_\perp>}}{L_\perp}w,
\end{equation}
so that the total volumetric heating rate due to the turbulent energy
dissipation equals:
\begin{equation}\label{eq:isotropic}
e=\frac1{L_\perp}\sqrt{\frac{w=w_-=w_+}\rho}(w_-+w_+).
\end{equation}
The introduction of $L_\perp\propto 1/k_\perp$, which is {\it the
transverse correlation length of turbulence} instead of
$\omega/V_A=k_\|$, which is the wave vector for the Alfven wave
propagating {\it along} the magnetic field line, makes the cascade theory
better suited to the nature of purely transverse Alfven waves. For such waves,
the ratio of the {\it transverse} turbulent velocity to the 
{\it longitudinal} wavelength could hardly characterize the rate 
of nonlinear wave dissipation. In more refined turbulent theory (which
we do not apply and even do not review here) the spectral energy should be
introduced separately for parallel and transverse wave
vectors. However, within our {\it phenomenological approach we assume that} 
$k_\perp\propto k_\|$ and combine, once needed, the spectral energy
distribution over $\omega=V_Ak_\|$ with the transverse corelation
length, to quantify the non-linear dissipation rate.

To close the model and to compare its predictions with the
unsigned-flux-based heating function, we chose, again following
\cite{hollw86}, the scaling law for the transverse correlation length:
\begin{equation}\label{eq:LPerp}
L_\perp\propto |{\bf B}|^{-1/2},\quad L_\perp=\frac{75\,[{\rm
    km}\cdot{\rm T}^{1/2}]}{\sqrt{|{\bf B}|}}.
\end{equation}
In the present work we use the following range of values for the
empirical constant: 
$L_\perp\sqrt{|{\bf B}|}$:
$$
50\,[{\rm km}\cdot{\rm T}^{1/2}]\le L_\perp\sqrt{|{\bf B}|}\le
100\,[{\rm km}\cdot{\rm T}^{1/2}]
$$
in the numerical estimates below we use $L_\perp\sqrt{|{\bf B}|}\approx
75\,[{\rm km}\cdot{\rm T}^{1/2}]$ from \cite{hollw86}. 
 The scaling law $|{\bf B}|L_\perp^2=const$
is well compatible with the ``percolation hypothesis'' noticed above:
while waves propagating from the solar interior along the flux tube
towards the solar surface it is natural to assume that
$L_\perp^2$ varies proportionally to the flux tube cross-section 

To discuss the value and the variability of the spatial scale in the
coronal heating, if the latter is related to the Alfven wave turbulence 
dissipation, compare Eqs.(\ref{eq:isotropic},\ref{eq:LPerp}) with the 
representation of the wave dissipation as in the right hand side of Eq.(\ref{eq:usma}). From this
comparison we find: $V_A/L=\sqrt{w/\rho}/L_\perp$. 
At the corona base we can employ Eq.(\ref{eq:w}) to evaluate the
boundary values of $w$ and the estimate for $L$ in this region reads:
$$
L=\frac{V_A(L_\perp \sqrt{|{\bf B}|})}{\sqrt{\frac{|{\bf
        B}|\left(\frac{P}{|{\bf
          B}|}\right)\sqrt{\mu_0\rho}}{\rho}}}
\approx\sqrt{\frac{V_A}{400\,{\rm km/s}}}40\,{\rm Mm}.
$$
We see that with the choice of the above estimates for the transverse
correlation length and for the boundary condition for the Poynting flux
the estimate of the dissipation length at the coronal base has only
weak scaling (as a square root) with the Alfven speed. The latter  does not vary too much at the
coronal base, as long as $V_A\propto |{\bf B}|/\sqrt{\rho}$ and the
density correlates with the magnetic field intensity, being larger in
the active regions with strong magnetic fields. We also see that for a
reasonable estimate for the Alfven speed at the coronal base,
$V_A\sim(200-500)$ km/s, the
dissipation length is close to the {\it ad hoc} value $L=40$ Mm as used in
\cite{downs10}. The dissipation length is not used below directly, as
long as Eqs.(\ref{eq:isotropic},\ref{eq:LPerp}) do not include
it. However, the capability to reproduce the popular unsigned-flux-based heating
model with the popular envelop function is important. We can now
strengthen the conclusion above in the following manner: {\it the choice
  of a boundary condition for the Poynting flux together with the
  estimates of the non-linear dissipation in an isotropic Kolmogorov
  turbulence allows us to reproduce in detail the unsigned-flux-based
  ad hoc heating model for the closed field region}.  
 
\subsection{Coronal Holes and the Solar Wind Model}
An important {\it ad hoc} approach while applying the heating function
is the capability to use strongly different functions in the coronal
holes and in the closed field regions. In \cite{downs10,lion09} in the
coronal holes the following heating function was applied:
$$e=5\cdot10^{-7}\exp[-(R-R_\odot)/(0.7R_\odot)]\,{\rm erg/cm^3s}.$$ 
Comparing
with the unsigned-flux-based model for the closed field region, the
spatial scale of the heating function in the coronal holes is 
significantly longer than $\sim 40$ Mm which we used above. 

Here, we discuss the possibility of implementing such a heating model via the Alfven
wave absorption. First, we do not assume that in the coronal holes the
Poynting flux at the corona base is different from that in the closed
field region and attribute the drastic difference in the heating
functions to the difference in the dissipation rate only. Second,
comparing the integral of the heating function over the coronal hole
volume with the accepted value of the Poynting flux for the turbulence, 
one can find that only a few percents of the Poynting flux is absorbed 
in the coronal holes within the heliocentric distance range. $R_{\odot}\le R\le 2.5
R_{\odot}$. 

{\it Comparing with the closed field region, where almost all
wave energy is absorbed, one can conclude that the wave energy
dissipation rate in coronal holes should include a small factor of
the order of a few percents compared with
Eqs.(\ref{eq:isotropic},\ref{eq:LPerp})}.
\subsubsection{Imbalanced Kolmogorov Turbulence and its Dissipation}
The drastic decrease in the wave dissipation rate in the coronal holes
comes naturally if we take into account a strong turbulence anisotropy
in the coronal holes: the outward propagating waves should be much
more intense than the waves propagating toward the Sun. Below, having in mind that the {\it anisotropy} is a natural property of any MHD turbulence in the 
directional magnetic field, the turbulence in the coronal holes is referred to as {\it imbalanced}, not anisotropic. 

Now we consider an imbalanced turbulence, such that $I_+\ne I_-$. As
long as the real cascade physics requires the presence of oppositely
propagating waves, we require that the {\it wave-wave interaction rate
  should be the function of minimum of $w_-,w_+$ in a strongly
imbalanced turbulence}. To
achieve this, instead of Eq.(\ref{eq:isotropic}) we use the following 
expression for the dissipation rate:
\begin{equation}\label{eq:imbalanced}
e=\frac{\sqrt{|{\bf B}|}}{(L_\perp\sqrt{|{\bf B}|})}
\left(\sqrt{\frac{w_+}{\rho}}w_-+\sqrt{\frac{w_-}{\rho}}w_+\right).
\end{equation}
For the closed field region this expression is more realistic than that
we used before, as long it correctly captures more intense heating
in shorter loops, in which the intensities of the oppositely
propagating waves are more uniform, as compared
to longer loops, in which the wave energy for counter-propagating waves
may be strongly non-uniform and imbalanced. However, for coronal
holes this approximation is not sufficient as long as the WKB
approximation, under which Eq.(\ref{gen}) was derived is not accurate enough. Within the model we use here, in the
coronal holes only outward propagating waves can exist. while in
reality the wave outward propagation is accompanied with some
reflection, resulting in appearance of waves propagating towards the
Sun. Small ratio of the reflected wave amplitude to the bulk wave amplitude
is described by the reflection coefficient, $C_{\rm refl}$. Therefore,
the neglibly small $\min(w_\pm)$ in the expression for the dissipation rate 
should be floored with
$C^2_{\rm refl}\max(w_\pm)$, giving an ultimate expression for the turbulent
dissipation rate we use in the present work:
\begin{equation}\label{eq:opposite}
e=\frac{\sqrt{|{\bf B}|}}{(L_\perp\sqrt{|{\bf B}|})}
\left(\sqrt{\frac{\max(w_+,C^2_{\rm refl}w_-)}{\rho}}w_-+\sqrt{\frac{\max(w_-,C^2_{\rm refl}w_+)}{\rho}}w_+\right).
\end{equation}
\subsubsection{How to Derive Eq.(\ref{eq:imbalanced})?}
In order to justify the Eqs.(\ref{eq:imbalanced},\ref{eq:opposite}), we outline 
the way to derive them consistently. One can employ the framework of reduced 
MHD, which solves the equations of motion, induction and continuity:
$$
\frac{\partial{\bf u}}{\partial t}+({\bf u}\cdot\nabla){\bf u}+
\frac{\nabla B^2}{\mu_u\rho}+\frac{\nabla(P_e+P_i)}\rho=\frac{({\bf B}\cdot\nabla){\bf B}}{\mu_0\rho},
$$
$$
\frac{\partial {\bf B}}{\partial t}+({\bf u}\cdot\nabla){\bf B}+{\bf B}(\nabla\cdot{\bf u})=({\bf B}\cdot\nabla){\bf u},
$$
$$
\frac{\partial \rho}{\partial t}+\nabla\cdot(\rho{\bf u})=0,
$$
by means of representing the magnetic field and velocity vectors as sums of 
regular and turbulent parts, ${\bf u}=\tilde{\bf u}+\delta{\bf u},\quad{\bf B}=
\tilde{\bf B}+\delta{\bf B}$ (below tildes are omitted) and by simplifying the 
equations for turbulent amplitudes:
\begin{equation}\label{eq:deltau}
\frac{\partial\delta{\bf u}}{\partial t}+({\bf u}\cdot\nabla)\delta{\bf u}+(\delta{\bf u}\cdot\nabla)\delta{\bf u}+(\delta{\bf u}\cdot\nabla){\bf u}
=\frac{({\bf B}\cdot\nabla)\delta{\bf B}}{\mu_0\rho}+\frac{(\delta{\bf B}\cdot\nabla)\delta{\bf B}}{\mu_0\rho}+\frac{(\delta{\bf B}\cdot\nabla){\bf B}}{\mu_0\rho},
\end{equation}
\begin{equation}\label{eq:deltaB}
\frac{\partial \delta{\bf B}}{\partial t}+({\bf u}\cdot\nabla)\delta{\bf B}+
(\delta{\bf u}\cdot\nabla)\delta{\bf B}+(\delta{\bf u}\cdot\nabla){\bf B}+\delta{\bf B}(\nabla\cdot{\bf u})=({\bf B}\cdot\nabla)\delta{\bf u}+(\delta{\bf B}\cdot\nabla)\delta{\bf u}+(\delta{\bf B}\cdot\nabla){\bf u},
\end{equation}
\begin{equation}\label{eq:deltarho}
\frac{\partial \rho}{\partial t}+({\bf u}\cdot\nabla)\rho+(\delta{\bf u}\cdot\nabla)\rho
+\rho\nabla\cdot{\bf u}=0,
\end{equation}
by assuming the incompressibility conditions: $\nabla\cdot\delta{\bf u}=0,\quad
{\bf B}\cdot\delta{\bf B}=0$. The equations for the Elsasser variables, 
${\bf z}_\pm =\delta{\bf u}\mp\delta{\bf B}/\sqrt{\mu_0\rho}$ are obtained as 
the sum Eq.(\ref{eq:deltau})$\mp\frac1{\sqrt{\mu_0\rho}}\times$Eq.(\ref{eq:deltaB})$\pm\frac{\delta{\bf B}}{\rho\sqrt{\mu_0\rho}}\times$Eq.(\ref{eq:deltarho}):
$$
\frac{d_\pm{\bf z}_\pm}{dt}
+{\bf z}_\mp\cdot\nabla{\bf u}\mp
\frac{{\bf z}_\mp\cdot\nabla{\bf B}}{\sqrt{\mu_0\rho}}-
\frac{{\bf z}_\pm-{\bf z}_\mp}{4\rho}
\frac{d_\mp\rho}{dt}
=0,
$$
where $\frac{d_\pm}{dt}=\frac\partial{\partial t}+({\bf u}\pm{\bf V}_A+{\bf z}_\mp)\cdot\nabla$ and ${\bf V}_A=\frac{\bf B}{\sqrt{\mu_0\rho}}$.  
The dynamic equations for the wave energy densities, $w_\pm=\rho{\bf z}^2_\pm/4$,
may be obtained by multypling the above equation by $\rho{\bf z}_\pm/2$ and 
adding Eq.(\ref{eq:deltarho})$\times3{\bf z}^2_\pm/8$:
$$
\frac{\partial w_\pm}{\partial t}+\nabla\cdot[({\bf u}\pm{\bf V}_A+{\bf z}_\mp)w_\pm)+\frac{w_\pm}2(\nabla\cdot{\bf u})+\frac\rho2{\bf z}_\pm\cdot[({\bf z}_\mp\cdot\nabla){\bf u}\mp\frac{({\bf z}_\mp\cdot\nabla){\bf B}}{\sqrt{\mu_0\rho}}]+\frac{{\bf z}_+\cdot{\bf z}_-}8\frac{d_\mp\rho}{dt}=0. 
$$ 
As the first (WKB) approximation one can set ${\bf z}_\mp=0$ in the equations 
for $w_\pm$ and obtain:
$$
\frac{\partial w_\pm}{\partial t}+\nabla\cdot[({\bf u}\pm{\bf V}_A)w_\pm)+\frac{w_\pm}2(\nabla\cdot{\bf u})=0.
$$
This equation describes Alfven wave propagation along the magnetic field lines 
(firs two terms) and the wave energy reduction in the expanding plasma 
(the last term) because of the work done by the wave pressure $P_w=\frac12(w_++w_-)$. Within the more accurate approximation the non-linear term $\nabla\cdot({\bf z}_\mp w_\pm)$ results in the turbulent cascade and the wave energy 
dissipation. The dissipation rate for the wave energy density, $w_+$, is 
controlled by the amplitude of the oppositely propagating wave, $|{\bf z}_-|\sim\sqrt{w_-/\rho}$, 
and the correlation length, $L_\perp$, in the transverse (with respect to the magnetic field) 
direction, 
because in the Alfven wave, propagating along the magnetic field, $\delta{\bf u},\,\delta{\bf B}$ 
are perpendicular to the magnetic field. Therefore, 
$\nabla\cdot({\bf z}_\mp w_\pm)\sim\frac1{L_\perp}\sqrt{\frac{w_\mp}\rho}w_\pm$ and the WKB equations 
with an account for non-linear dissipation read:
$$
\frac{\partial w_+}{\partial t}+\nabla\cdot[({\bf u}+{\bf V}_A)w_+)+\frac{w_+}2(\nabla\cdot{\bf u})
=-\frac1{L_\perp}\sqrt{\frac{w_-}\rho}w_+,
$$
$$
\frac{\partial w_-}{\partial t}+\nabla\cdot[({\bf u}-{\bf V}_A)w_-)+\frac{w_-}2(\nabla\cdot{\bf u})=-\frac1{L_\perp}\sqrt{\frac{w_+}\rho}w_-.
$$
These equations work both for the balanced and moderately imbalanced turbulence and for 
the balanced turbulence ($w_+=w_-=w$) they reduce to the equations by \cite{hollw86} as described 
above. For the imbalanced turbulence they properly reduce the dissipation rate for the dominant 
wave by expressing this rate in terms the amplitude of the minor oppositely 
propagating wave.

However, the above equations for ${\bf z}_\pm$ demonstrate that even in a linear
approximation the WKB approach dismisses the correlations between inward and outward 
propagating waves. Indeed, in the partial differential equation for ${\bf z}_+$ there are the source terms present linearly proportional to ${\bf z}_-$ and vise versa, while in the WKB 
approximation these source terms describing the conversion between the oppositely going 
waves are omitted. The omitted correlations are important (see \cite{tu95,dmit02}) and we will include 
them into the model as described in the forthcoming publication. 

Here, we parameterize this effect in order to incorporate the coronal heating in the coronal holes. The WKB 
approximation predicts no inward propagating waves originating from the open magnetic field lines, while in reality 
the inward propagating wave should arise from the partial conversion (``{\it reflection}'') of the outward 
propagating wave due to non-WKB effects. {\it The non-WKB generation of the inward propagating waves is 
parameterized via their amplitude related to that of the outward propagating waves, so that the maximum of the WKB 
and non-WKB wave amplitudes is used to determine the dissipation rate of the dominant outward propagating waves:}      
\begin{equation}\label{eq:wplusfull}
\frac{\partial w_+}{\partial t}+\nabla\cdot[({\bf u}+{\bf V}_A)w_+]+\frac{w_+}2(\nabla\cdot{\bf u})
=-\frac1{L_\perp}\sqrt{\frac{\max(w_-,C^2_{\rm refl}w_+)}\rho}w_+,
\end{equation}
\begin{equation}\label{eq:wminusfull}
\frac{\partial w_-}{\partial t}+\nabla\cdot[({\bf u}-{\bf V}_A)w_-)+\frac{w_-}2(\nabla\cdot{\bf u})=-\frac1{L_\perp}\sqrt{\frac{\max(w_+,C^2_{\rm refl}w_-)}\rho}w_-.
\end{equation}
\subsubsection{Estimates for the Reflection Coefficient}
In the present research we focus on the study of the Lower Corona and
do not compare the solar wind generation and its propagation towards 1
AU. Therefore, for our present purpose the estimate 
$$
C_{\rm refl}=const\sim0.01\div0.1
$$
is sufficient.

Having in mind to develop this approach in the forthcoming
publications, we discuss briefly, how the idea of the WSA semi-empirical model
can be implemented within the Alfv\'en-wave-turbulence-driven model. As 
discussed above, for the WSA model the speed of the solar wind
originating from the given magnetic field line footpoint may be found
from two characteristics of the line - the expansion factor,
$f_{\rm exp}$, and the distance to the coronal hole boundary, $\theta$:
$$
u_\infty=u_\infty(\theta,f_{\rm exp}).
$$
In the original version of the WSA model, for example, the solar wind speed,
was a function of the expansion factor only:
$u_\infty=A/f_{\rm exp}^\delta$, where the unknown constants, $A$ and 
$\delta$ were chosen for better fitting the observed solar wind speed
at 1 AU. 

Once formulated in terms of the Alfv\'en wave absorption, the model can
no longer employ the {\it existing WSA formulae} for the solar wind, as
long as there is no simple relationship between the wave absorption and
the solar wind speed. However, we can keep using {\it the idea of the
  WSA model} and fit the observed solar wind parameters at 1 AU by
properly choosing a formula for $C_{\rm refl}$ that depends on
$\theta$ and $f_{\rm exp}$. The first choice to be tested is:
\begin{equation}
C_{\rm refl}=A\cdot f_{\rm exp}^\delta,
\end{equation} 
where the unknown constants will be chosen to better fit the
observations.

Note that there is some reasoning in favor of this approach within the
wave-based model for the solar wind, as long as  the expansion
factor for a given magnetic field line is a good characteristic of the 
reflection coefficient for the waves, propagating along this
line. Indeed, the expansion factor is larger for strongly bent
magnetic field lines, but the reflection coefficients at such lines is
also larger than on straight lines. At small distances from the coronal
boundary the abrupt gradients of the plasma density also can be a
reason for the increase in the reflection coefficient (see \cite{evan12}), 
resulting in the generally recognized opinion that the slow slow wind (higher wave 
absorption) has its origin near the coronal hole boundary (small
$\theta$).
\section{Computational Model}
\label{CT}

In this section, we describe the governing equations to be solved
numerically as well as the numerical tools we use in the numerical 
simulations.
\subsection{Governing Equations}
The model includes the standard MHD equations (non-specified
denotations are as usually): 
\begin{equation}\label{eq:cont}
\frac{\partial\rho}{\partial t}+\nabla\cdot(\rho{\bf u})=0,
\end{equation}
\begin{equation}\label{eq:induction}
\frac{\partial{\bf B}}{\partial t}+\nabla\cdot\left({\bf u}\otimes{\bf
  B}-{\bf B}\otimes{\bf u}\right)=0,
\end{equation}
\begin{equation}\label{eq:momentum}
\frac{\partial(\rho{\bf u})}{\partial t}+\nabla\cdot\left(\rho{\bf
  u}\otimes{\bf u}-\frac{{\bf B}\otimes{\bf B}}{\mu_0}\right)+\nabla\left(P_i+P_e+\frac{B^2}{2\mu_0}+\frac{w_-+w_+}2\right)=0,
\end{equation}
with the full energy equations applied separately to ions 
\begin{eqnarray}\label{eq:energy}
\frac{\partial\left(\frac{P_i}{\gamma-1}+\frac{\rho u^2}2\right)}{\partial
  t}&+&\nabla\cdot\left[\left(\frac{\rho u^2}2+\frac{\gamma P_i}{\gamma-1}+\frac{B^2}{\mu_0}\right){\bf
  u}-\frac{{\bf B}({\bf u}\cdot{\bf B})}{\mu_0}\right]=\nonumber\\
&=& -({\bf u}\cdot\nabla)P_e+
\frac{N_ik_B}{\tau_{ei}}\left(T_e-T_i\right)+\Gamma_-w_-+\Gamma_+w_+,
\end{eqnarray}
and to electrons:
\begin{equation}\label{eq:electron}
\frac{\partial\left(\frac{P_e}{\gamma-1}\right)}{\partial
  t}+\nabla\cdot\left(\frac{P_e}{\gamma-1}{\bf
  u}\right)+P_e\nabla\cdot{\bf u}=\nabla\left(\kappa\nabla 
T_e\right)+\frac{N_ik_B}{\tau_{ei}}\left(T_i-T_e\right)-Q_{\rm rad}.
\end{equation}
In addition to the standard effects, the above equations account for a
possible difference in the electron and ion temperatures, the electron
heat conduction parallel to the magnetic field lines:
\begin{equation}
\kappa=\frac{{\bf B}\otimes{\bf B}}{B^2}\kappa_\|,\quad \kappa_\|\propto T_e^{5/2}
\end{equation}
the radiation energy loss from an optically thin plasma:
\begin{equation}
Q_{\rm rad}=N_eN_i\Lambda(T_e),
\end{equation}
as well as the energy exchange between electron and ions
parameterized via the relaxation time, $\tau_{ei}$, as this is usually
done. The Alfven wave turbulence pressure and dissipation rate is
applied in the above equations. In particular simulations, the heating
due to the turbulence dissipation (see Eq.\ref{eq:energy}) may be split
between electron and ions and we can also use the total energy equation
for electron and ions, to improve the computational efficiency. The
turbulence propagation and dissipation are described within the WKB approximation: 
\begin{equation}\label{eq:wave}
\frac{\partial w_{\pm}}{\partial t}+\nabla\cdot\left({\bf
  u}w_{\pm}\pm{\bf b}V_Aw_{\pm}\right)+\frac12w_{\pm}\nabla\cdot{\bf
  u}=-\Gamma_\pm w_{\pm}.
\end{equation}
The total wave energy dissipation (that is the total of the right hand side of 
Eqs.(\ref{eq:wave}) taken with the opposite signs), $e=\Gamma_- w_-+\Gamma_+ w_+$, is included in the right hand side of the energy equation (\ref{eq:energy}) as 
the source term.
Summarize the above consideration and derivations regarding the
dissipation rate (see Eqs.(\ref{eq:opposite},\ref{eq:wplusfull},\ref{eq:wminusfull})), which we apply in the following form:
\begin{equation}\label{eq:absorption}
\Gamma_\pm=\frac{\sqrt{|{\bf B}|}}{(L_\perp\cdot\sqrt{|{\bf
      B}|})}\sqrt{\frac{\max(w_\mp,C^2_{\rm refl}w_\pm)}{\rho}},
\end{equation}
and with the following ranges for the parameters involved:
\begin{equation}\label{eq:LPerpSI}
(L_\perp\cdot\sqrt{|{\bf
      B}|})=(20\div100)\,{\rm km\cdot T^{1/2}},\quad C_{\rm refl}=0.01\div0.1,
\end{equation}
We use the simplest equation of state for the coronal plasma with the
polytropic index, $\gamma=\frac53$. 

The system of governing equations is solved numerically using
BATS-R-US/SWMF code (see section 3.3 below). The boundary condition for the
Poynting flux (or for the intensity of the outgoing waves) is given by
Eq.(\ref{eq:bc}) (see Fig.~1). The boundary condition for the coronal magnetic field
is taken from the full disk magnetogram. The boundary condition for the
density and temperature requires more attention and is discussed in the
following subsection.
\begin{figure}\label{fig0}
\plottwo{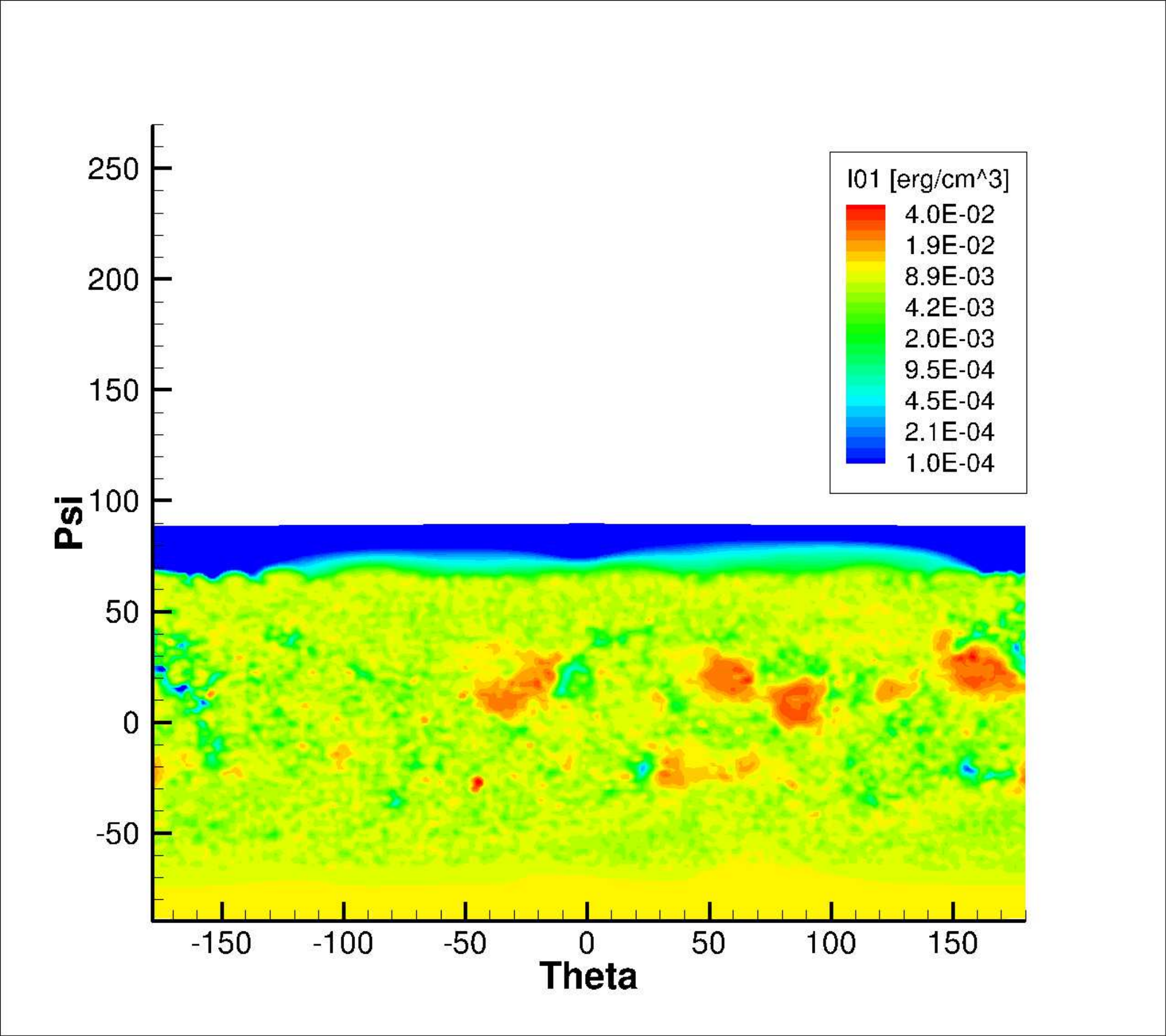}{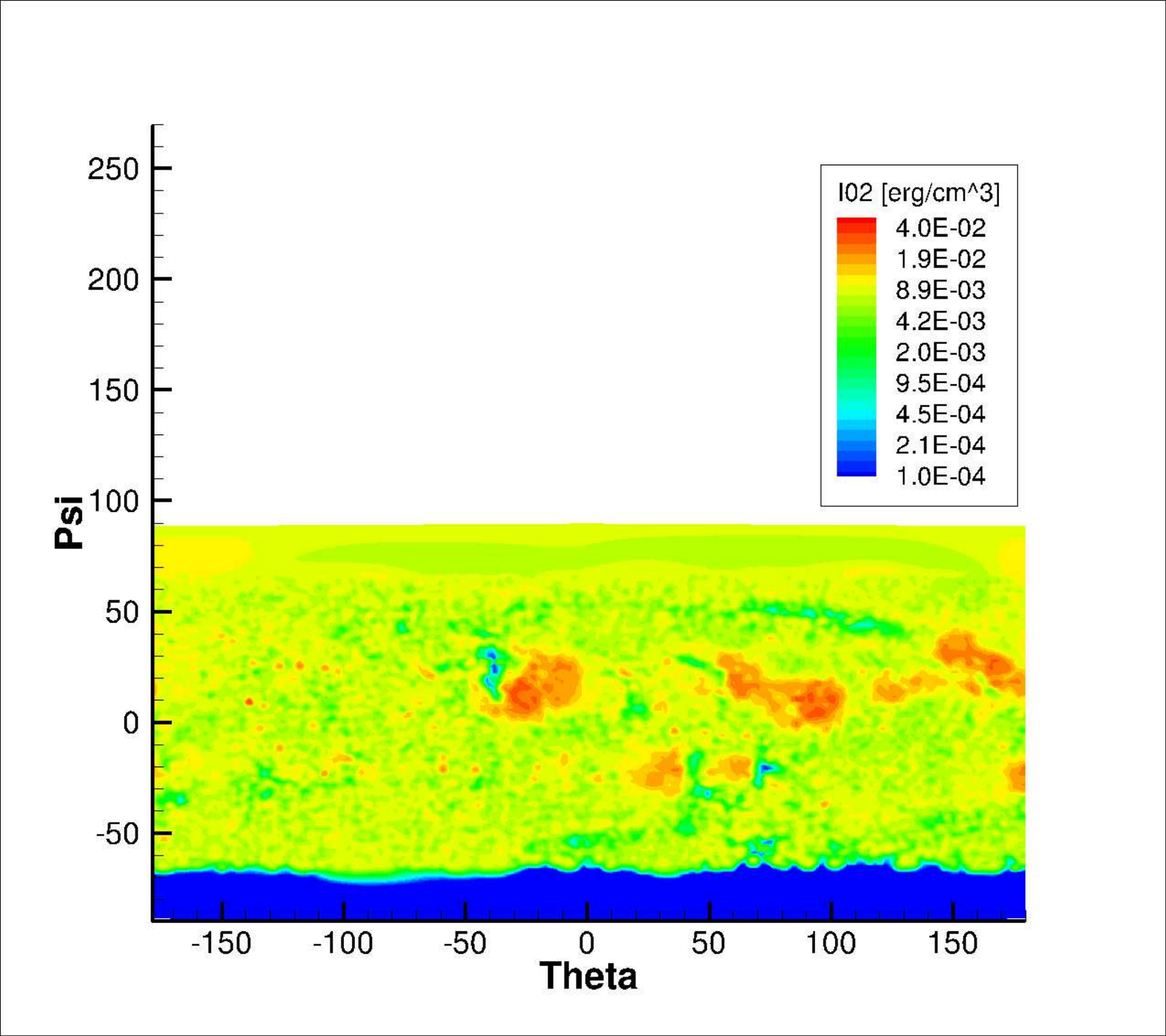}
\caption{The distributions of the wave energy densities, $w_+$ 
(left panel), and $w_-$ (right panel), at the heliospheric distance of 
$1.003\,R_\odot$ illustrate boundary contitions for the Poynting flux.}
\end{figure}
\subsection{Chromosphere and transition region}
\subsubsection{Chromosphere Boundary Condition}
Here, we discuss the analytical solution of the hydrostatic equilibrium
and heat transfer equations with the exponential heating function, 
$Q_h=A\exp(-x/L)$, The solution is as follows:
$$
T_e=T_i=T_0,\qquad
N_e=N_i=N_0
\exp\left(
-\frac{m_igx}{k_B(T_e+T_i)}\right),
$$
\begin{equation}
Q_h=Q_{rad}=N_e^2\Lambda(T_0)=N_0^2\Lambda(T_0)
\exp\left(-\frac{m_igx}{k_BT_0}\right).
\end{equation}
Here $g=274\ {\rm m/s^2}$ is the gravity acceleration near the solar surface,
the direction of this acceleration being antiparallel to the x-axis and
$m_i$ is the proton mass. The two constants in the solution,
$N_0$ and $T_0$, which are the boundary values for the density and
temperature correspondingly, are unambiguously related to the
amplitude and decay length of the heating function:
\begin{equation}
L=\frac{k_BT_0}{m_ig}\approx T_0\cdot(30\ {\rm m/K}),\qquad A=N_0^2\Lambda(T_0).
\end{equation}  
Notice a very simple relationship for the exponential decay length for
the heating function, which is the half of the barometric scale length
for density: $2L=L_g=k_B(T_e+T_i)/(m_ig)$. 

The described solution satisfies the equation for the heat conduction
as long as the heat transfer in the isothermic solution is absent and
heating at each point exactly balances the radiation cooling. The
hydrostatic equilibrium is also maintained, as long as
$$
k_B\frac{\partial(N_eT_e+N_iT_i)}{\partial x}=-gN_im_i.
$$ 

The suggested solution well describes the chromosphere. The short
decay length of the heating function, which is equal to $\approx0.6$ Mm
for $T_0=2\cdot10^4$ K may be, presumably, related to absorption of
(magneto)acoustic turbulent waves, rapidly damping due to the
wave-breaking effects, which result in the shock wave formation and energy dissipation.    
Physically, including this chromosphere heating 
function would imply that the temperature in the chromosphere is elevated 
compared to the photospheric temperatures due to some mechanism acting
in the chromosphere itself. By no means can this energy be transported
from the solar corona as long as the electron heat conduction rate at
chromospheric temperatures is very low.  

As long as we do not apply such {\it ad hoc} short-scale heating
function in the chromosphere, nor we include short-scale turbulent
dissipation length, {\it we set the boundary condition for the density and
the temperature at the top of chromosphere}:
\begin{equation}\label{eq:trhobc}
T_{ch}=T_0=2\cdot 10^4\,{\rm K},\qquad N_{ch}=N_0\approx2\cdot
10^{16}\,{\rm m^{-3}}
\end{equation} 
\subsubsection{Transition region}
The analytical solution for the transition region had been published
many times. Here we focus on the following issues: (1) merging this
solution to that for the chromosphere; (2) the validity of zero-gravity 
approximation; and (3) the validity of the modified heat conduction 
model.

The heat transfer equation for a steady state hydrogen plasma in 
a uniform magnetic field reads:
\begin{equation}\label{transition}
\frac\partial{\partial s}\left(\kappa_0T_e^{5/2}\frac{\partial
  T_e}{\partial s}\right) +Q_h-N_e^2\Lambda(T_e)=0.
\end{equation}
Here $Q_h=\Gamma(w_-+w_+)$ is the coronal heating function assumed to be constant at 
the spatial scales typical for the transition region. 
Note that the coordinate is taken along the
magnetic field line, not along the radial direction.

On multiplying Eq.(\ref{transition}) by $\kappa_0T_e^{5/2}(\partial
T_e/\partial s)$ and by integrating from the interface to chromosphere
till a given point at a temperature, $T_e$, one can obtain:
\begin{equation}\label{transition1}
[
\frac12
\kappa_0^2 T_e^5
\left(
\frac{\partial T_e}{\partial s}
\right)^2
+\frac27\kappa_0 Q_h T_e^{7/2}
]|^{T_e}_{T_{ch}}=
(N_eT_e)^2
\int^{T_e}_{T_{ch}}{\kappa_0T^{1/2}\Lambda(T)dT}.
\end{equation}
Here the product, $N_eT_e$, is assumed to be constant, therefore, it is
separated from the integrand. 
For a given $T_{ch}$ the only parameter in the solution is
$(N_eT_e)$. In can be expressed at any point in terms of the local
value of the heating flux and the radiation loss integral:
\begin{equation}\label{transition2}
(N_eT_e)=\sqrt{\frac{\frac12
\kappa_0^2 T_e^5
\left(
\frac{\partial T_e}{\partial s}
\right)^2
+\frac27\left(\kappa_0 Q_h T_e^{7/2}-\kappa_0 Q_hT_{ch}^{7/2}\right)}{\int^{T_e}_{T_{ch}}{\kappa_0T^{1/2}\Lambda(T)dT}}}.
\end{equation} 

{\it The assumption} of constant $(N_eT_e)$ is fulfilled only if the effect
of gravity is negligible. Quantitatively this condition is not 
trivial, as long as both the barometric scale and especially the heat
conduction scale are the functions of temperature. The barometric scale
may be approximated as $L_g(T_e)\approx T_e\cdot(60\ {\rm m/K}) $. The heat conduction scale, $L_h$, can be
estimated by noticing that within the large part of the transition
region the radiation losses dominate over the heating function,
therefore, they are balanced by heat conduction: $\kappa_0
T_e^{5/2}\cdot(Te/L_h^2)\sim Q_r$. Thus, the condition for
neglecting gravity is:
\begin{equation}\label{restriction}
L_g(T_e)\approx T_e\cdot(60\ 
{\rm m/K})\gg 
L_h\approx
\sqrt{ \frac{\kappa_0T_e^{9/2}}{\Lambda(T_e)(N_eT_e)^2}}.
\end{equation}
\begin{figure}\label{fig_length}
\includegraphics[scale=0.9]{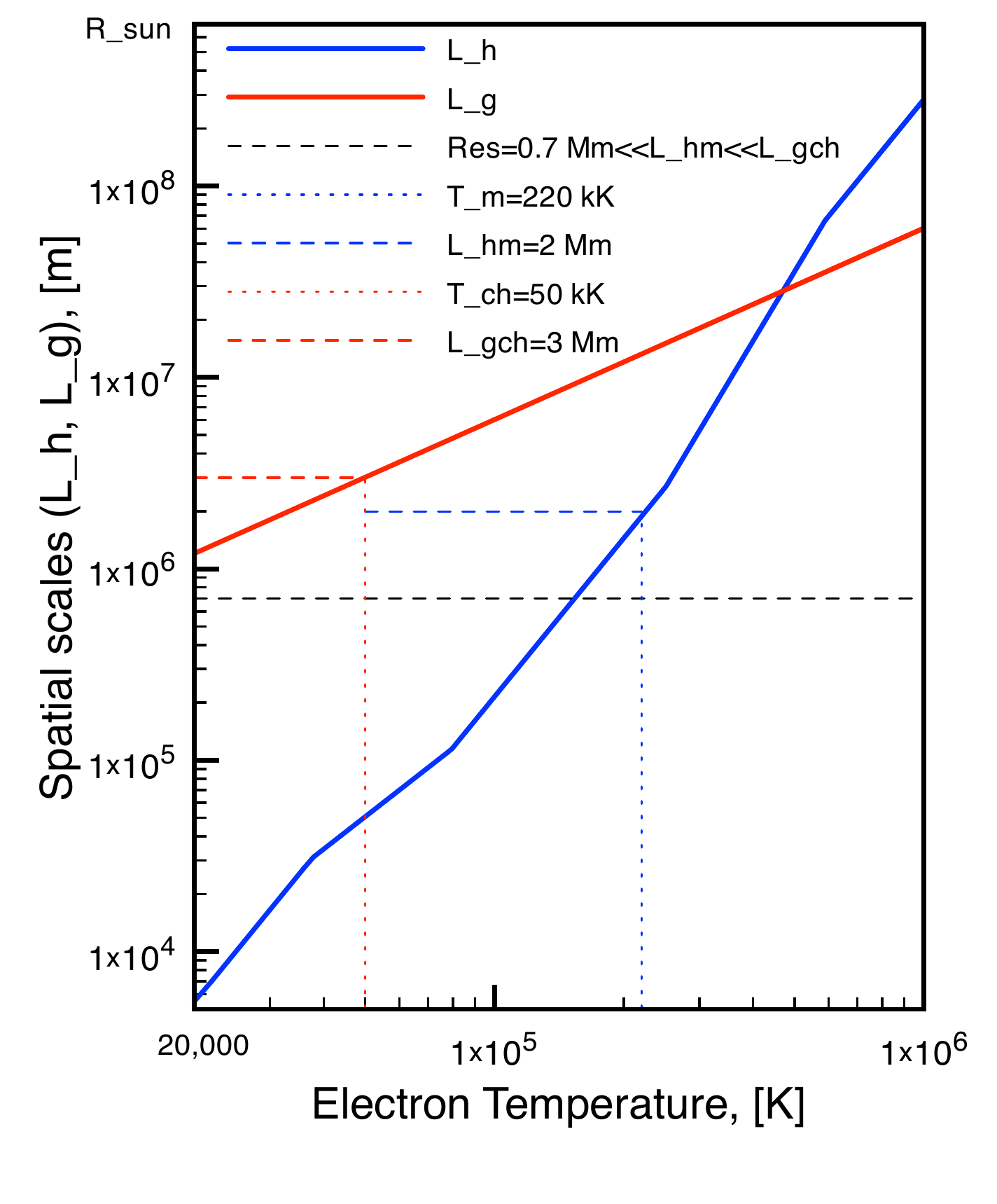}
\caption{Typical scales of the transition region: the heat
conduction scale (blue), $L_h$, and the gravitational height (red), $L_g$, as 
functions of the temperature. The real transition region (at $T_e\le 4\cdot10^5$ K) is very narrow 
comparing to the gravitational height: $L_h\ll L_g$. To keep this property in 
simulations we chose the chromosphere temperature, $T_{ch}=5\cdot10^4$ K, and 
the temperature below which to modify the electron heat conduction, 
$T_m=2.2\cdot10^5$ K, in such a way that $\Delta x<L_h<L_g$ at 
$T_{ch}\le T\le T_{m}$ with the transition region artificially extended till 
$L_h(T_m)\approx2$ Mm. Here $\Delta x=10^{-3}R_\odot=0.7$ Mm is the spatial 
resolution of our grid near the Sun.}
\end{figure} 
In Fig.~2 we plot temperature dependencies, $L_h(T_e),\ L_g(T_e)$, for
$(N_eT_e)=10^{20}\ {\rm K/m^3}$. We see that near the
chromosphere boundary the approximation (\ref{restriction}) works very well as long as the
temperature changes with height are very abrupt. The increase in
temperature till $10^5{\rm K}$ occurs at the length shorter than 0.1 Mm. This estimate agrees with the temperature profile
recovered from observations of chromospheric lines (see, e.g., Fig.2 and Fig.8
in \cite{avre08}).  However, as the temperature further grows with 
the height, the gravity effect on the tmeperature and density profiles
becomes more significant. It becomes comparable with the heat
conduction effect at $T_e\approx 4.5\cdot 10^5$ K, which can accepted
as the coronal base temperature, 
so that  the
transition region 
corresponds to the temperature
range from $T_{ch}\approx 2\cdot 10^4$ K till $4.5\cdot10^5$ K, with a
typical width being $\sim 10\ {\rm Mm}\approx R_S/70$. 
The transition
region solution merges to the chromoshere solution with negligible jump in 
pressure. The merging point in the chromosphere, therefore, is at 
the density of $(N_eT_e)/T_{ch}\sim 10^{16}\ {\rm m^{-3}}$. The
 short heat conduction scale at the chromosphere temperature
(see Fig1) ensures that the heat flux from the solar corona across the
transition region does not penetrate towards higher densities.

It is known that Eq.(\ref{transition2}) may be used to establish the
boundary condition for density at the 'coronal base',
however, we use another 
way to model the transition region by increasing artificially
the heat conduction in the lower temperature range (see \cite{abbe07}).  
Consider a following transformation of the temperature functions in  
Eqs.(\ref{transition}-\ref{transition1}):
\begin{equation}
\kappa_0\rightarrow f\kappa_0,\qquad ds\rightarrow f\ ds,\qquad,
\Gamma\rightarrow\Gamma/f,\qquad Q_{rad}\rightarrow Q_{rad}/f,
\end{equation}
with a common factor, $f\ge1$. The equations does not change in this 
transformation and the only effect on the solution is that the
temperature profile in the transition region becomes a factor of $f$
wider. By applying the factor, $f=(T_m/T_e)^{5/2}$ at $T_{ch}\le T_e\le
T_m$, one can achieve that the heat conduction scale in this range is
almost constant and is close to $\approx 2$ Mm for a choice of
$T_m\approx 2.2\cdot 10^5$ K (see Fig.~1).

It should be emphasized, however, that the choice of the temperature
range to apply this transformation is highly restricted by the
condition as in (\ref{restriction}). In choosing a higher value of
$T_m$ the heat conduction scale at the chromospheric temperature
exceeds the barometric scale in the chromosphere resulting in
physically meaningless penetration of the coronal heat to the deeper
chromosphere. The global model of the solar corona with this
unphysical energy sink suffer from the reduced values of the coronal 
temperature and produces a visible distortion in the EUV and X-ray
synthetic images. {\it Thus, in formulating the transition region model
we modify the heat conduction, the radiation loss rate and the wave
dissipation rate and the maximal temperature for this modification does
not exceed $T_m\approx 2.2\cdot 10^5$ K.}

\subsection{BATS-R-US and SWMF Codes}
\label{CT_BATSRUS}

The BATS-R-US (Block Adaptive Tree Solar Wind Roe-type Upwind Scheme)
code has been developed at the UM.  It solves the equations of ideal
MHD---a system of eight equations describing the transport of mass,
momentum, energy, and magnetic flux \cite[]{groth00,powell99}.  This
massively parallel code enables Sun-to-Earth simulations to be performed
in near real-time when run on hundreds of processors on a supercomputer
\cite[]{manchester04b}.  The implementation of adaptive-mesh-refinement
(AMR) in BATS-R-US allows orders of magnitude variation in numerical
resolution within the computational domain.  This is important for a global
model of the solar magnetic fields in which one strives to resolve such
structures as shocks, volumetric currents of flux ropes, electric current
sheets in a 3D domain, which may extend out to hundreds of $R_\odot$.
The use of AMR also enables us to resolve the fine structure of active
regions on the Sun, which spawn CMEs.  This is vitally necessary for
incorporating high-resolution magnetic data into a numerical MHD model.
In the context of solar-heliospheric physics, BATS-R-US has been utilized
to model the global structure of the solar corona and solar wind
\cite[]{roussev03b,cohen07,cohen08}, as well as the initiation
\cite[]{roussev03a,jacobs09} and propagation of idealized
\cite[]{manchester04a, manchester04b} and not-so-idealized
\cite[]{roussev04,roussev07} solar eruptions and associated SEP
events \cite[]{sokolov04}. 

The SWMF (Space Weather Modeling Framework) is a high-performance
computational tool that has been developed at the University of Michigan
to simulate the coupled Sun-Earth system \cite[]{toth04,toth05}. One
of the main modules within the SWMF is the BATS-R-US MHD code.
The SWMF is a structured collection of software building blocks to
develop Sun-Earth system modeling components, in order to couple
and assemble them into applications.  The framework was designed
to have ``plug-and-play'' capabilities, and presently it links together nine
inter-operating models of physics domains, ranging from the surface of
the Sun to the upper atmosphere of the Earth.  Tying these models
together gives a self-consistent whole in which each region is described
by a world-class model, and those models communicate data with each
other.  The SWMF has achieved faster than real-time performance on
massively parallel computers, such as the NASA's Columbia supercluster.

\section{Simulation Results for CR2107 and Comparison with
  Observations}

At the rising phase of solar cycle 24, the Sun is becoming more and more active. In this study, we focus on the Carrington Rotation 2107 (From 2011 February 16 to 2011 March 16). During this carrington rotation, a series of interesting events happened on March 7 in NOAA 11164. First, an M3.7 flare occurred around 20:00 UT, followed by a very fast CME with speed $\sim$2200 km s$^{-1}$. A gradual SEP event was observed at 1 AU, which suggests that the CME-driven shock may play an important role to accelerate particles. Also, gamma-ray emission above 100 MeV was detected by Fermi/LAT and lasted for $\sim$8 hr. This is very unusual since the hours-long gamma ray emission from the Sun was observed only three times in the past. Therefore, the flare-related acceleration is extremely strong in this event. By simulating this event and validating the results with the observations, we can achieve a better understanding of the physics behind the varies phenomenon from the Sun to 1 AU.

\subsection{Steady State Solar Wind and Validation}
To simulate the CR2107, a data-driven boundary condition is used for
the inner boundary magnetic field. We use the SDO/HMI synoptic 
magnetogram in this study (see Fig.~3). Since the uncertainty of
  the polar region field, a correction is made to the polar field
  \citep{sun11}. To extrapolate the initial potential field, Finite
  Difference Iterative Potential-field Solver (FDIPS) is used
  \citep{toth11}. The resolution of the magnetogram is
  3600$\times$1440.
\begin{figure}
\includegraphics[scale=0.6]{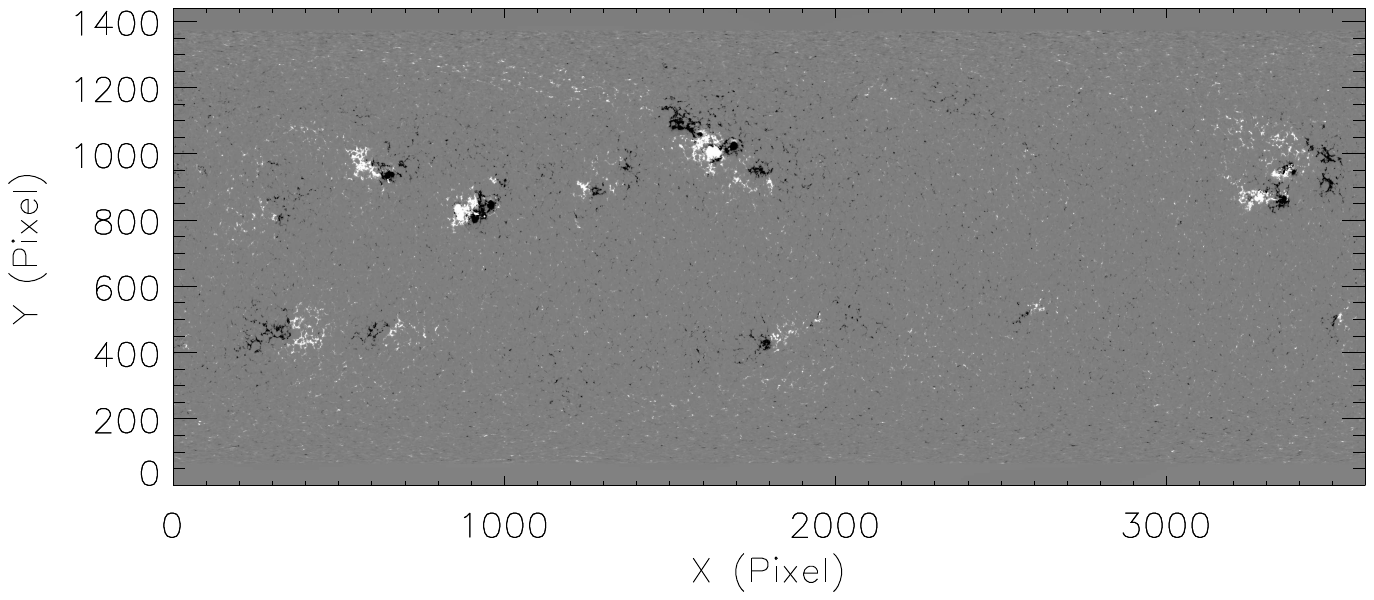}
\caption{The synoptic magnetogram of Carrington Rotation 2107 from SDO/HMI. The saturation value of the magnetic field is set to be 200 Gauss in order to show the active regions more clearly.}
\end{figure}

A spherical grid is used in the simulation. The simulation domain reaches 20 R$_{\odot}$. In total, 1.3$\times$10$^{5}$ blocks (6*4*4) is used with 1.2$\times$10$^{7}$ cells. Adaptive mesh refinement is performed to resolve the heliosphere current sheet (HCS).  The smallest cell reaches $\sim$10$^{-3}$ R$_{\odot}$, while the largest is $\sim$0.8 R$_{\odot}$. The radial velocity at X=0 plane is shown in Fig.~4. The fast and slow winds are evident in that plot. The fast wind is $\sim$600 km s$^{-1}$. The slow wind is $\sim$300 km s$^{-1}$.

In Figs.~5-6, zoom-in velocity and plasma beta with magnetic field lines are shown near the Sun. The velocity pattern shows more complex structures comparing with the solar minimum condition. The magnetic field show many structures (e.g., streams, pseudo-streams). From the plot of the plasma beta, we can see the different regions of the corona have very different plasma beta. The  polar regions with open field lines have small beta $\sim$0.01 near the Sun and increase outward. The high beta regions show the location of the current sheets in which the magnetic field is very small.

In order to evaluate the new steady state solar wind model near the
Sun, we compare the model output to the 
electron temperature and density derived from EUV images of the Sun by
using the DEMT method \citep{fraz10}. In general, the DEMT
method uses a time series of EUV images 
under the assumption of no time variation and uniform solar rotation to
derive 3D emissivity distribution in 
each EUV band. By Local Differential Emission Measure (LDEM) analysis,
the 3D distribution of the electron 
density and temperature can be obtained. The DEMT method assumes the
plasma is optically thin. In this study, 
we use three bands of EUV observation (171, 193, and 284 $A$) from SDO/AIA.

In Figs.~7-8, we show the ratio between the model and DEMT output for the electron density and temperature. The sphere is at r=1.05R$_{\odot}$. The ring is between r=1.035R$_{\odot}$ and r=1.225R$_{\odot}$. For the density, we see a good agreement between the model and the DEMT output for most regions. The active regions and polar regions in the model have relatively smaller density (Note that the DEMT method has larger uncertainty for the active and polar regions). For temperature, we find the model and DEMT have better agreement for higher altitude. The model temperature at  r=1.05R$_{\odot}$ is lower than the DEMT by a factor of $\sim$2.

In Fig.9 we provide a direct comparison of the observed EUV images with
those synthesized from simulations. For better visibility we marked the
active regions and coronal holes. One can see that our numerical
simulations well reproduce the observed morphological structures, thus
confirming the physical reasonings of the new global coronal model.
\section{Conclusion and future work}
At the present stage the global coronal model which is the heart of the
solar-heliosphere model in the SWMF does not employ any longer the
{\it ad hoc} heating function, with no significant loss in the results
quality. For the Carrington rotation CR2107 the simulation results are
compared with observations. The uncertainty ranges in the model are
comparatively narrow, except for the reflection coefficient and the
uncertainties may be further reduced in the course of more thorough
validation with observations. The contrast ratio in the synthetic
images will also be improved.

Another direction for our further investigation is to improve the solar
wind model in the way described in sub-subsection 2.4.2. The reflection
coefficient at the time is the most uncertain parameter and the only
point which is certain about the realistic reflection is that it should
account for the wave interaction with realistic gradients of the
magnetic field and the coronal density. The solar wind propagation to
1AU and the results comparison with the {\it in situ} observations will
be presented in the accompanying publications.  

\begin{figure}
\includegraphics[scale=0.50]{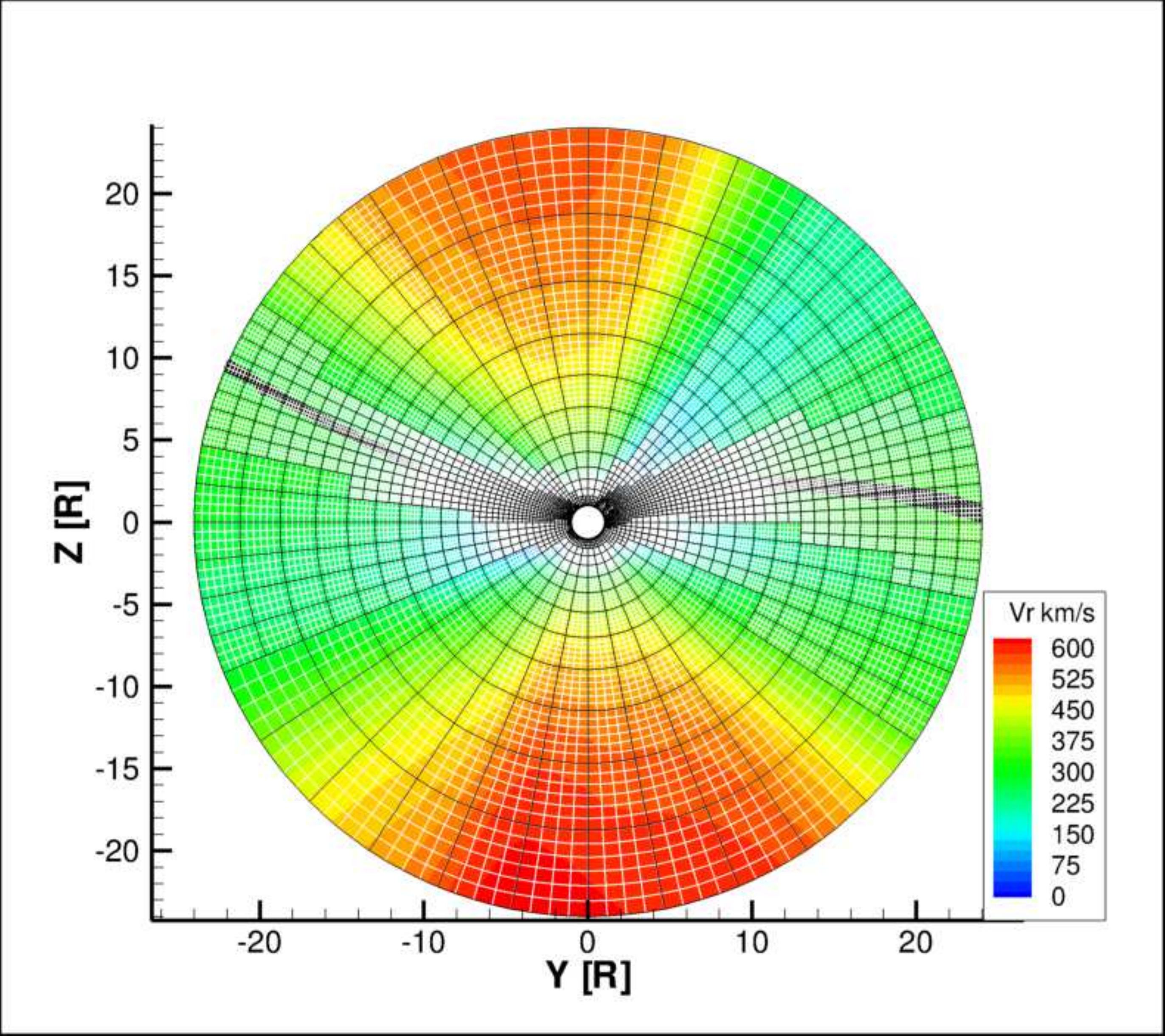}
\caption{The radial velocity of the steady state solar wind at X=0 plane. The black boxes in the Figure show the blocks and the white boxes show the cells. The thick black line near the equator shows the location of the heliospheric current sheet.}
\end{figure}

\begin{figure}
\includegraphics[scale=0.50]{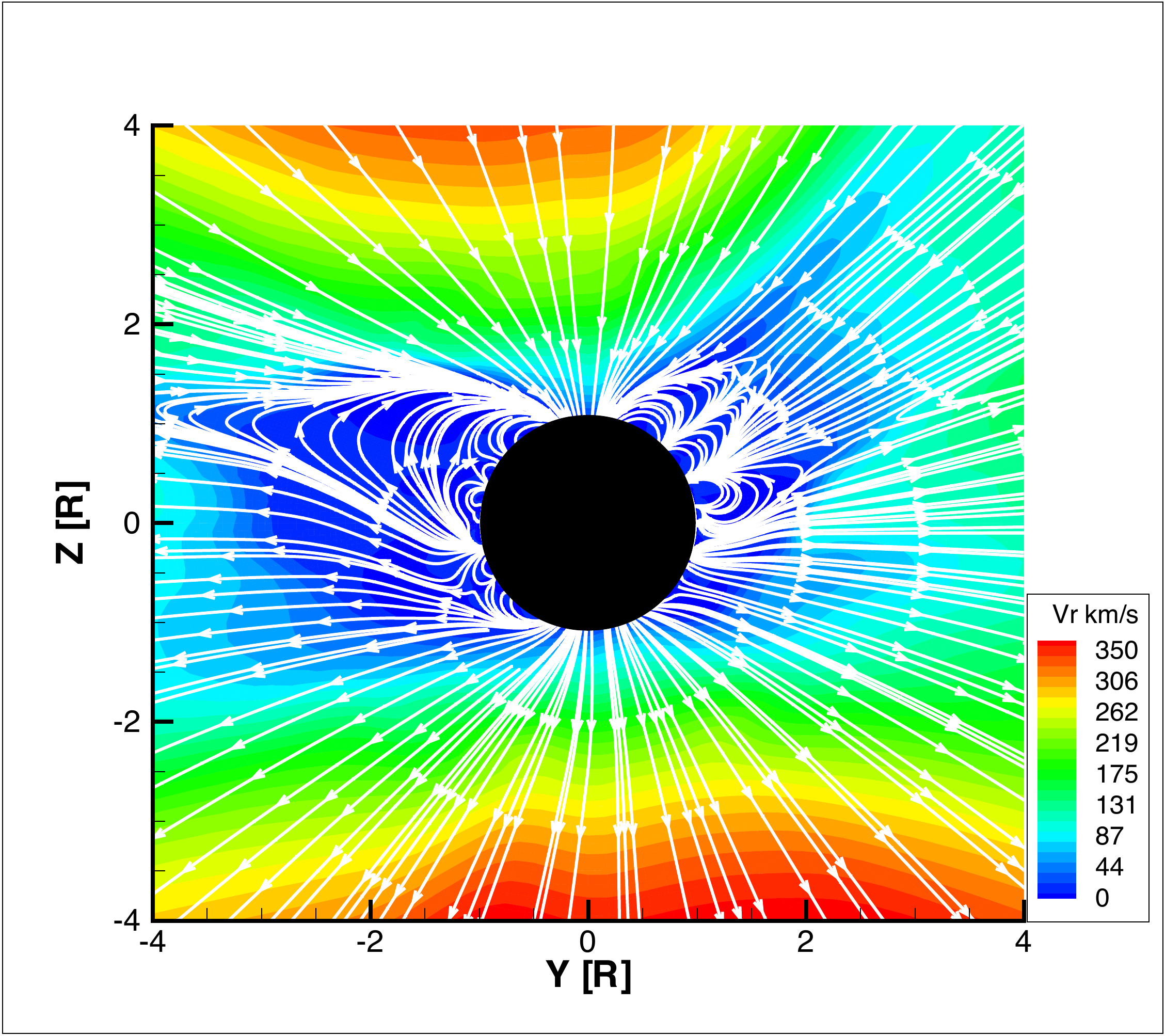}
\caption{The radial velocity field with magnetic field lines at X=0 plane from -4 R$_{\odot}$ to 4 R$_{\odot}$.}
\end{figure}

\begin{figure}
\includegraphics[scale=0.50]{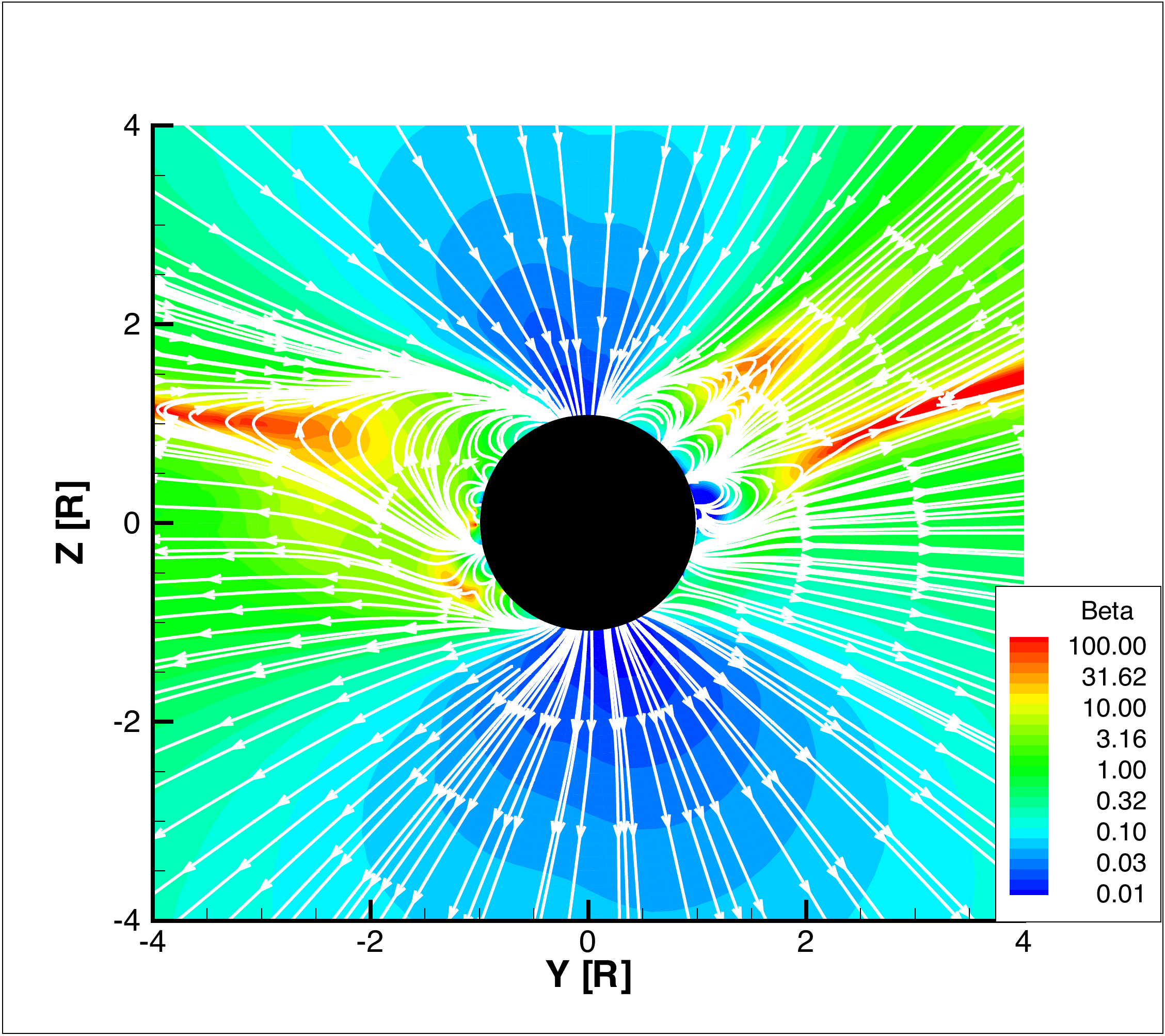}
\caption{The plasma beta with magnetic field lines at X=0 plane from -4 R$_{\odot}$ to 4 R$_{\odot}$..}
\end{figure}

\begin{figure}
\includegraphics[scale=0.45]{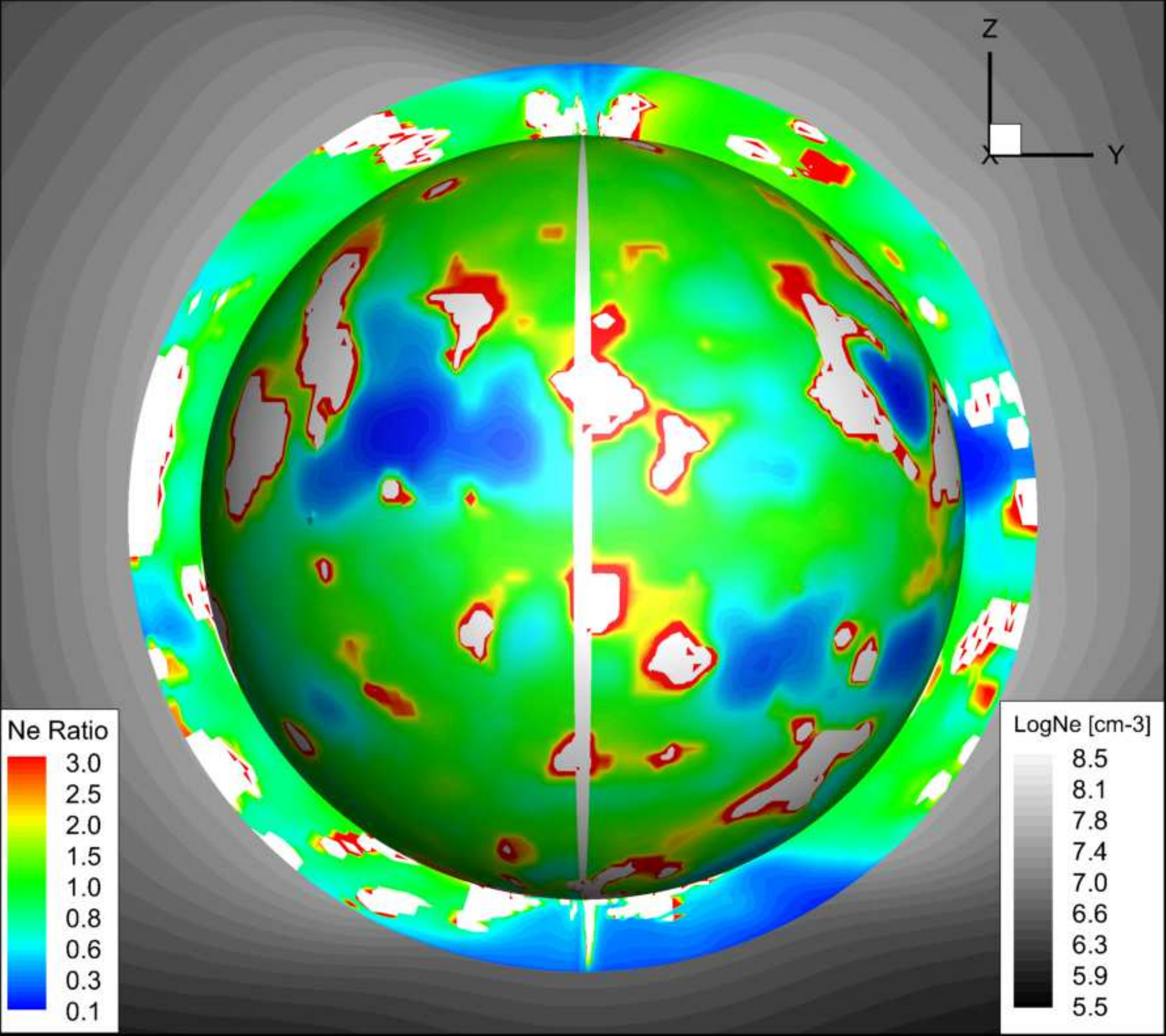}
\caption{Comparison between the SC and DEMT output near the Sun for electron number density. The inner ring shows the ratio between the model and DEMT output from 1.035 $R_{\odot}$ to 1.225 $R_{\odot}$ and outside background shows the model output. The iso-surface of the Sun shows the same ratio at R = 1.05 $R_{\odot}$.}
\end{figure}

\begin{figure}
\includegraphics[scale=0.45]{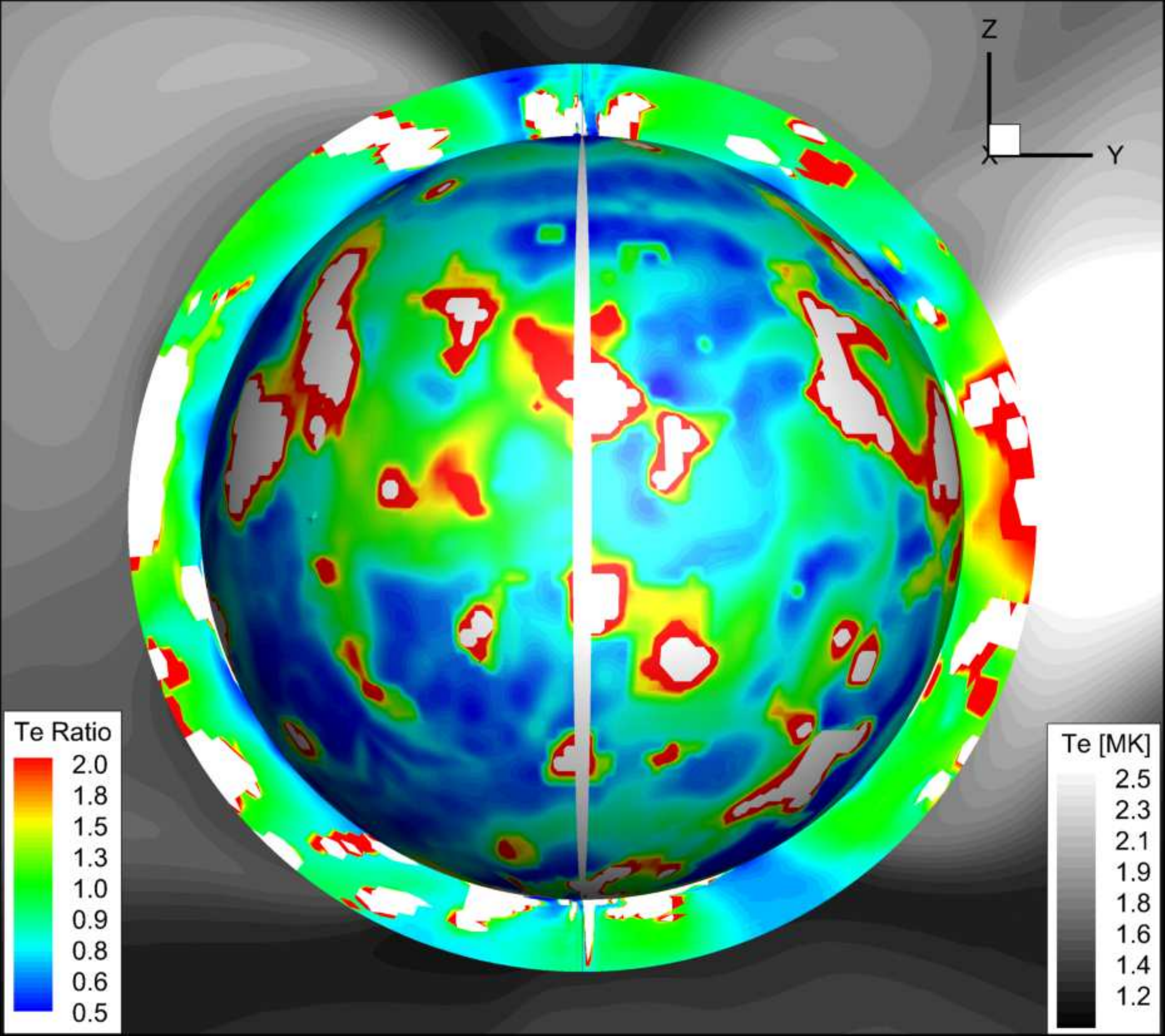}
\caption{Comparison between the SC and DEMT output near the Sun for electron temperature. The inner ring shows the ratio between the model and DEMT output from 1.035 $R_{\odot}$ to 1.225 $R_{\odot}$ and outside background shows the model output. The iso-surface of the Sun shows the same ratio at R = 1.05 $R_{\odot}$.}
\end{figure}

\begin{figure}
\includegraphics[scale=0.6]{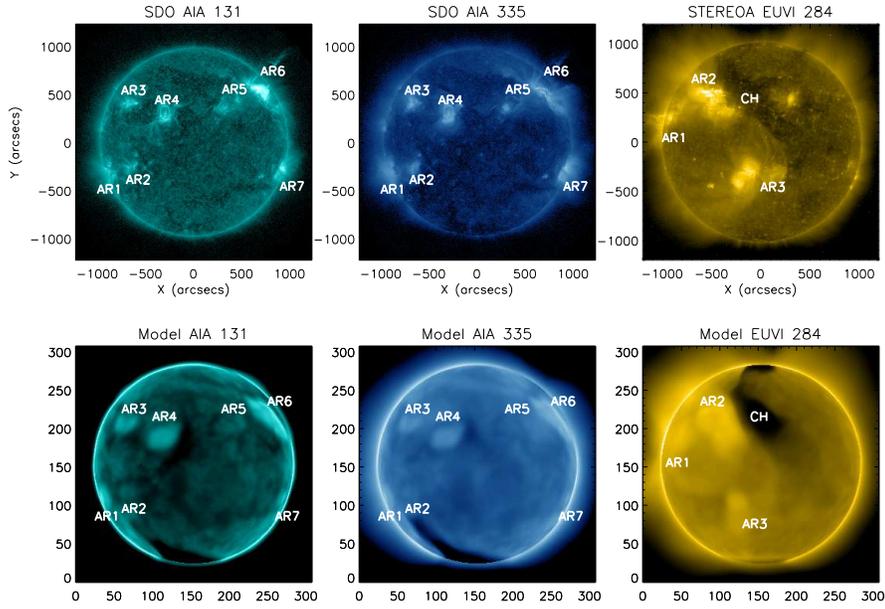}
\caption{The comparison between observations and synthesized EUV images
  of the model. Top panels (left to right): Observational images from SDO AIA 131
  {\AA}, SDO AIA 335 {\AA}, and STEREO A EUVI 284 {\AA}. The
  observation time is 2011 March 7 20:00 UT. Bottom panels:
  synthesized EUV images of the model. The active regions and coronal
  hole are marked both in the observational and synthesized images, to
  demonstrate the good reproducibility of the observed morphological
  structures in our simulations.}
\end{figure}

\clearpage

\addcontentsline{toc}{section}{References}

\bibliographystyle{authyear}
\bibliography{wholeilr}

\begin{thebibliography}{{\em {Suzuki} and {Inutsuka}}(2005}

\bibitem[{\em {Abbett}}(2007)]{abbe07}
{Abbett}, W.~P.,
\newblock {The Magnetic Connection between the Convection Zone and Corona in
  the Quiet Sun},
\newblock {\em Astrophys. J.}, {\em 665}, 1469--1488, August 2007.

\bibitem[{\em {Arge} and {Pizzo}}(2000)]{arge00}
{Arge}, C.~N., and V.~J. {Pizzo},
\newblock {Improvement in the Prediction of Solar Wind Conditions Using
  Near-Real Time Solar Magnetic Field Updates},
\newblock {\em J. Geophys. Res.}, {\em 105}, 10,465--10,480, May 2000.

\bibitem[{\em {Avrett} and {Loeser}}(2008)]{avre08}
{Avrett}, E.~H., and R.~{Loeser},
\newblock {C},
\newblock {\em Astrophys. J. Suppl.}, {\em 175}, 229, June 2008.

\bibitem[{\em {Cohen} et~al.}(2007)]{cohen07}
{Cohen}, O., I.~V. {Sokolov}, I.~I. {Roussev}, C.~N. {Arge}, W.~B.
  {Manchester}, T.~I. {Gombosi}, R.~A. {Frazin}, H.~{Park}, M.~D. {Butala},
  F.~{Kamalabadi}, and M.~{Velli},
\newblock {A Semiempirical Magnetohydrodynamical Model of the Solar Wind},
\newblock {\em Astrophys. J. Lett.}, {\em 654}, L163--L166, January 2007.

\bibitem[{\em {Cohen} et~al.}(2008)]{cohen08}
{Cohen}, O., I.~V. {Sokolov}, I.~I. {Roussev}, and T.~I. {Gombosi},
\newblock {Validation of a Synoptic Solar Wind Model},
\newblock {\em J. Geophys. Res. (Space Physics)}, {\em 113}( A12 ), A03104,
  March 2008.

\bibitem[{\em {Cranmer}}(2010)]{cran10}
{Cranmer}, S.~R.,
\newblock {An Efficient Approximation of the Coronal Heating Rate for use in
  Global Sun-Heliosphere Simulations},
\newblock {\em Astrophys. J.}, {\em 710}, 676--688, February 2010.

\bibitem[{\em {De Pontieu} et~al.}(2007)]{dupo08}
{De Pontieu}, B., S.~W. {McIntosh}, M.~{Carlsson}, V.~H. {Hansteen}, T.~D.
  {Tarbell}, C.~J. {Schrijver}, A.~M. {Title}, R.~A. {Shine}, S.~{Tsuneta},
  Y.~{Katsukawa}, K.~{Ichimoto}, Y.~{Suematsu}, T.~{Shimizu}, and S.~{Nagata},
\newblock {Chromospheric Alfv{\'e}nic Waves Strong Enough to Power the Solar
  Wind},
\newblock {\em Science}, {\em 318}, 1574--, December 2007.

\bibitem[{\em {Dmitruk} et~al.}(2002)]{dmit02}
{Dmitruk}, P., W.~H. {Matthaeus}, L.~J. {Milano}, S.~{Oughton}, G.~P. {Zank},
  and D.~J. {Mullan},
\newblock {Coronal Heating Distribution Due to Low-Frequency, Wave-driven
  Turbulence},
\newblock {\em \apj}, {\em 575}, 571--577, August 2002.

\bibitem[{\em {Downs} et~al.}(2010)]{downs10}
{Downs}, C., I.~I. {Roussev}, B.~{van der Holst}, N.~{Lugaz}, I.~V. {Sokolov},
  and T.~I. {Gombosi},
\newblock {Toward a Realistic Thermodynamic Magnetohydrodynamic Model of the
  Global Solar Corona},
\newblock {\em Astrophys. J.}, {\em 712}, 1,219--1,231, April 2010.

\bibitem[{\em {Evans} et~al.}(2012)]{evan12}
{Evans}, R.~M., M.~{Opher}, R.~{Oran}, B.~{van der Holst}, I.~V. {Sokolov},
  R.~{Frazin}, T.~I. {Gombosi}, and A.~{V{\'a}squez},
\newblock {Coronal Heating by Surface Alfv{\'e}n Wave Damping: Implementation
  in a Global Magnetohydrodynamics Model of the Solar Wind},
\newblock {\em \apj}, {\em 756}, 155, September 2012.

\bibitem[{\em {Farrugia} et~al.}(1997)]{farr97}
{Farrugia}, C.~J., V.~A. {Osherovich}, and L.~F. {Burlaga},
\newblock {The non-linear evolution of magnetic flux ropes: 3. effects of
  dissipation},
\newblock {\em Annales Geophysicae}, {\em 15}, 152--164, February 1997.

\bibitem[{\em {Fisk} and {Schwadron}}(2001)]{fisk01b}
{Fisk}, L.~A., and N.~A. {Schwadron},
\newblock {The Behavior of the Open Magnetic Field of the Sun},
\newblock {\em Astrophys. J.}, {\em 560}, 425--438, October 2001.

\bibitem[{\em {Fisk} et~al.}(1999a)]{fisk99c}
{Fisk}, L.~A., N.~A. {Schwadron}, and T.~H. {Zurbuchen},
\newblock {Acceleration of the Fast Solar Wind by the Emergence of New Magnetic
  Flux},
\newblock {\em J. Geophys. Res.}, {\em 104}, 19,765--19,772, September 1999a.

\bibitem[{\em {Fisk} et~al.}(1999b)]{fisk99b}
{Fisk}, L.~A., T.~H. {Zurbuchen}, and N.~A. {Schwadron},
\newblock {On the Coronal Magnetic Field: Consequences of Large-Scale Motions},
\newblock {\em Astrophys. J.}, {\em 521}, 868--877, August 1999b.

\bibitem[{\em {Fisk}}(1996)]{fisk96}
{Fisk}, L.~A.,
\newblock {Motion of the Footpoints of Heliospheric Magnetic Field Lines at the
  Sun: Implications for Recurrent Energetic Particle Events at High
  Heliographic Latitudes},
\newblock {\em J. Geophys. Res.}, {\em 101(A7)}, 15,547--15,554, July 1996.

\bibitem[{\em {Fisk}}(2001)]{fisk01a}
{Fisk}, L.~A.,
\newblock {On the Global Structure of the Heliospheric Magnetic Field},
\newblock {\em J. Geophys. Res. (Space Physics)}, {\em 106(A8)},
  15,849--15,858, August 2001.

\bibitem[{\em {Groth} et~al.}(2000)]{groth00}
{Groth}, C.~P.~T., D.~L. {DeZeeuw}, T.~I. {Gombosi}, and K.~G. {Powell},
\newblock {Global Three-Dimensional MHD Simulation of a Space Weather Event:
  CME Formation, Interplanetary Propagation, and Interaction with the
  Magnetosphere},
\newblock {\em J. Geophys. Res.}, {\em 105}, 25,053--25,078, November 2000.

\bibitem[{\em {Hollweg}}(1986)]{hollw86}
{Hollweg}, J.~V.,
\newblock {Transition Region, Corona and Solar Wind in Coronal Holew},
\newblock {\em J. Geophys. Res.}, {\em 91}, 1411--1425, April 1986.

\bibitem[{\em {Hu} et~al.}(2000)]{hu00}
{Hu}, Y.~Q., R.~{Esser}, and S.~R. {Habbal},
\newblock {A four-fluid turbulence-driven solar wind model for preferential
  acceleration and heating of heavy ions},
\newblock {\em \jgr}, {\em 105}, 5093--5112, March 2000.

\bibitem[{\em {Jacobs} et~al.}(2009)]{jacobs09}
{Jacobs}, C., I.~I. {Roussev}, N.~{Lugaz}, and S.~{Poedts},
\newblock {The Internal Structure of Coronal Mass Ejections: Are all Regular
  Magnetic Clouds Flux Ropes?},
\newblock {\em Astrophys. J. Lett.}, {\em 695}, L171--L175, April 2009.

\bibitem[{\em {Jin} et~al.}(2012)]{jin12}
{Jin}, M., B.~{Manchester}, W. B.and~{van der Holst}, J.~R. {Gruesbeck}, R.~A.
  {Frazin}, E.~{Landi}, A.~M. {Vasquez}, P.~L. { Lamy}, A.~{Llebaria},
  A.~{Fedorov}, G.~{Toth}, and T.~I. {Gombosi},
\newblock A global two-temperature corona and inner heliosphere model: a
  comprehensive validation study,
\newblock {\em \apj}, {\em 754}, 6, Mar 2012.

\bibitem[{\em {Li} and {Habbal}}(2003)]{Habb03}
{Li}, X., and S.~R. {Habbal},
\newblock {Coronal Loops Heated by Turbulence-driven Alfv{\'e}n Waves},
\newblock {\em \apjl}, {\em 598}, L125--L128, December 2003.

\bibitem[{\em {Li} et~al.}(2011)]{li11}
{Li}, G., B.~{Miao}, Q.~{Hu}, and G.~{Qin},
\newblock Effect of current sheets on the solar wind magnetic field power
  spectrum from the ulysses observation: From kraichnan to kolmogorov scaling,
\newblock {\em Phys. Rev. Lett.}, {\em 106}, 125001, Mar 2011.

\bibitem[{\em {Lionello} et~al.}(2001)]{lion01}
{Lionello}, R., J.~A. {Linker}, and Z.~{Miki{\' c}},
\newblock {Including the Transition Region in Models of the Large-Scale Solar
  Corona},
\newblock {\em Astrophys. J.}, {\em 546}, 542--551, January 2001.

\bibitem[{\em {Lionello} et~al.}(2009)]{lion09}
{Lionello}, R., J.~A. {Linker}, and Z.~{Miki{\'c}},
\newblock {Multispectral Emission of the Sun During the First Whole Sun Month:
  Magnetohydrodynamic Simulations},
\newblock {\em Astrophys. J.}, {\em 690}, 902--912, January 2009.

\bibitem[{\em {Manchester} et~al.}(2004a)]{manchester04a}
{Manchester}, W.~B., T.~I. {Gombosi}, I.~{Roussev}, D.~L. {De Zeeuw}, I.~V.
  {Sokolov}, K.~G. {Powell}, G.~{T{\' o}th}, and M.~{Opher},
\newblock {Three-Dimensional MHD Simulation of a Flux Rope Driven CME},
\newblock {\em J. Geophys. Res.}, {\em 109(A18)}, 1,102--1,119, January 2004a.

\bibitem[{\em {Manchester} et~al.}(2004b)]{manchester04b}
{Manchester}, W.~B., T.~I. {Gombosi}, I.~{Roussev}, A.~{Ridley}, D.~L. {De
  Zeeuw}, I.~V. {Sokolov}, K.~G. {Powell}, and G.~{T{\' o}th},
\newblock {Modeling a Space Weather Event from the Sun to the Earth: CME
  Generation and Interplanetary Propagation},
\newblock {\em J. Geophys. Res.}, {\em 109(A18)}, 2,107--2,122, February 2004b.

\bibitem[{\em {McIntosh} et~al.}(2011)]{mcintosh11}
{McIntosh}, S.~W., B.~{de Pontieu}, M.~{Carlsson}, V.~{Hansteen}, P.~{Boerner},
  and M.~{Goossens},
\newblock {Alfv{\'e}nic waves with sufficient energy to power the quiet solar
  corona and fast solar wind},
\newblock {\em Nature}, {\em 475}, 477--480, July 2011.

\bibitem[{\em Osman et~al.}(2011)]{osman11}
Osman, K.~T., W.~H. Matthaeus, A.~Greco, and S.~Servidio,
\newblock {Evidence for Inhomogeneous Heating in the Solar Wind},
\newblock {\em Astrophys. J. Lett.}, {\em {727}}( {1} ), January {2011}.

\bibitem[{\em {Pevtsov} et~al.}(2003)]{Pevtsov}
{Pevtsov}, A.~A., G.~H. {Fisher}, L.~W. {Acton}, D.~W. {Longcope}, C.~M.
  {Johns-Krull}, C.~C. {Kankelborg}, and T.~R. {Metcalf},
\newblock {The Relationship Between X-Ray Radiance and Magnetic Flux},
\newblock {\em Astrophys. J.}, {\em 598}, 1387--1391, December 2003.

\bibitem[{\em {Powell} et~al.}(1999)]{powell99}
{Powell}, K.~G., P.~L. {Roe}, T.~J. {Linde}, T.~I. {Gombosi}, and D.~L. {De
  Zeeuw},
\newblock {A Solution-Adaptive Upwind Scheme for Ideal Magnetohydrodynamics},
\newblock {\em J. Comp. Phys.}, {\em 154}, 284--309, September 1999.

\bibitem[{\em {Riley} et~al.}(2006)]{rile06}
{Riley}, P., J.~A. {Linker}, Z.~{Miki{\'c}}, R.~{Lionello}, S.~A. {Ledvina},
  and J.~G. {Luhmann},
\newblock {A Comparison between Global Solar Magnetohydrodynamic and Potential
  Field Source Surface Model Results},
\newblock {\em Astrophys. J.}, {\em 653}, 1510--1516, December 2006.

\bibitem[{\em {Roussev} et~al.}(2003a)]{roussev03a}
{Roussev}, I.~I., T.~G. {Forbes}, T.~I. {Gombosi}, I.~V. {Sokolov}, D.~L.
  {DeZeeuw}, and J.~{Birn},
\newblock {A Three-Dimensional Flux Rope Model for Coronal Mass Ejections Based
  on a Loss of Equilibrium},
\newblock {\em Astrophys. J. Lett.}, {\em 588}, L45--L48, May 2003a.

\bibitem[{\em {Roussev} et~al.}(2003b)]{roussev03b}
{Roussev}, I.~I., T.~I. {Gombosi}, I.~V. {Sokolov}, M.~{Velli},
  W.~{Manchester}, D.~L. {DeZeeuw}, P.~{Liewer}, G.~{T{\' o}th}, and
  J.~{Luhmann},
\newblock {A Three-Dimensional Model of the Solar Wind Incorporating Solar
  Magnetogram Observations},
\newblock {\em Astrophys. J. Lett.}, {\em 595}, L57--L61, September 2003b.

\bibitem[{\em {Roussev} et~al.}(2004)]{roussev04}
{Roussev}, I.~I., I.~V. {Sokolov}, T.~G. {Forbes}, T.~I. {Gombosi}, M.~A.
  {Lee}, and J.~I. {Sakai},
\newblock {A Numerical Model of a Coronal Mass Ejection: Shock Development with
  Implications for the Acceleration of GeV Protons},
\newblock {\em Astrophys. J. Lett.}, {\em 605}, L73--L76, April 2004.

\bibitem[{\em {Roussev} et~al.}(2007)]{roussev07}
{Roussev}, I.~I., N.~{Lugaz}, and I.~V. {Sokolov},
\newblock {New Physical Insight on the Changes in Magnetic Topology during
  Coronal Mass Ejections: Case Studies for the 2002 April 21 and August 24
  Events},
\newblock {\em Astrophys. J. Lett.}, {\em 668}, L87--L90, October 2007.

\bibitem[{\em {Sokolov} et~al.}(2004)]{sokolov04}
{Sokolov}, I.~V., I.~I. {Roussev}, T.~I. {Gombosi}, M.~A. {Lee}, J.~{K{\'
  o}ta}, T.~G. {Forbes}, W.~B. {Manchester}, and J.~I. {Sakai},
\newblock {A New Field-Line-Advection Model for Solar Particle Acceleration},
\newblock {\em Astrophys. J. Lett.}, {\em 616}, L171--L174, December 2004.

\bibitem[{\em {Sokolov} et~al.}(2009)]{sokolov09}
{Sokolov}, I.~V., I.~I. {Roussev}, M.~{Skender}, T.~I. {Gombosi}, and A.~V.
  {Usmanov},
\newblock {Transport Equation for MHD Turbulence: Application to Particle
  Acceleration at Interplanetary Shocks},
\newblock {\em Astrophys. J.}, {\em 696}, 261--267, May 2009.

\bibitem[{\em {Sun} et~al.}(2011)]{sun11}
{Sun}, X., Y.~{Liu}, J.~T. {Hoeksema}, K.~{Hayashi}, and X.~{Zhao},
\newblock { },
\newblock {\em Solar Physics}, {\em 270}, 9, September 2011.

\bibitem[{\em {Suzuki} and {Inutsuka}}(2005)]{suzu05}
{Suzuki}, T.~K., and S.-i. {Inutsuka},
\newblock {Making the Corona and the Fast Solar Wind: A Self-consistent
  Simulation for the Low-Frequency Alfv{\'e}n Waves from the Photosphere to 0.3
  AU},
\newblock {\em Astrophys. J. Lett.}, {\em 632}, L49--L52, October 2005.

\bibitem[{\em {Suzuki}}(2006)]{suzu06}
{Suzuki}, T.~K.,
\newblock {Forecasting Solar Wind Speeds},
\newblock {\em Astrophys. J. Lett.}, {\em 640}, L75--L78, March 2006.

\bibitem[{\em {Titov} et~al.}(2008)]{tito08}
{Titov}, V.~S., Z.~{Mikic}, J.~A. {Linker}, and R.~{Lionello},
\newblock {1997 May 12 Coronal Mass Ejection Event. I. A Simplified Model of
  the Preeruptive Magnetic Structure},
\newblock {\em Astrophys. J.}, {\em 675}, 1614--1628, March 2008.

\bibitem[{\em {T{\'o}th} et~al.}(2004)]{toth04}
{T{\'o}th}, G., O.~{Volberg}, A.~J. {Ridley}, T.~I. {Gombosi}, D.~L. {DeZeeuw},
  K.~C. {Hansen}, D.~{Chesney}, Q.~F. {Stout}, K.~G. {Powell}, K.~{Kane}, and
  R.~{Oehmke},
\newblock {A Physics-Based Software Framework for Sun-Earth Connection
  Modeling},
\newblock In {Lui}, A.~T.~Y., Y.~{Kamide}, and G.~{Consolini}, editors, {\em
  Proc. of the Conf. on the Sun-Earth Connection: Multi-scale Coupling of
  Sun-Earth Processes}. Elsevier Publ. Co.: Amsterdam, The Netherlands, 2004.

\bibitem[{\em {T{\'o}th} et~al.}(2005)]{toth05}
{T{\'o}th}, G., I.~V. {Sokolov}, T.~I. {Gombosi}, D.~R. {Chesney}, C.~R.
  {Clauer}, D.~L. {De Zeeuw}, K.~C. {Hansen}, K.~J. {Kane}, W.~B. {Manchester},
  R.~C. {Oehmke}, K.~G. {Powell}, A.~J. {Ridley}, I.~I. {Roussev}, Q.~F.
  {Stout}, O.~{Volberg}, R.~A. {Wolf}, S.~{Sazykin}, A.~{Chan}, B.~{Yu}, and
  J.~{K{\'o}ta},
\newblock {Space Weather Modeling Framework: A New Tool for the Space Science
  Community},
\newblock {\em J. Geophys. Res.}, {\em 110}, 12,226--12,237, December 2005.

\bibitem[{\em {T{\'o}th} et~al.}(2011)]{toth11}
{T{\'o}th}, G., B.~{van der Holst}, and Z.~{Huang},
\newblock { },
\newblock {\em Astrophys. J.}, {\em 732}, 102, September 2011.

\bibitem[{\em {Tu} and {Marsch}}(1995)]{tu95}
{Tu}, C.-Y., and E.~{Marsch},
\newblock {MHD structures, waves and turbulence in the solar wind: Observations
  and theories},
\newblock {\em \ssr}, {\em 73}, 1--210, July 1995.

\bibitem[{\em {Tu} and {Marsch}}(1997)]{tu97}
{Tu}, C.-Y., and E.~{Marsch},
\newblock {Two-Fluid Model for Heating of the Solar Corona and Acceleration of
  the Solar Wind by High-Frequency Alfven Waves},
\newblock {\em \solphys}, {\em 171}, 363--391, April 1997.

\bibitem[{\em {Usmanov} et~al.}(2000)]{usma00}
{Usmanov}, A.~V., M.~L. {Goldstein}, B.~P. {Besser}, and J.~M. {Fritzer},
\newblock {A Global MHD Solar Wind Model with WKB Alfv{\' e}n Waves: Comparison
  with Ulysses Data},
\newblock {\em J. Geophys. Res.}, {\em 105}, 12,675--12,696, June 2000.

\bibitem[{\em {V{\'a}squez} et~al.}(2010)]{fraz10}
{V{\'a}squez}, A.~M., R.~A. {Frazin}, and W.~B. {Manchester},
\newblock { },
\newblock {\em Astrophys. J.}, {\em 715}, 1352, September 2010.

\bibitem[{\em {Verdini} et~al.}(2010)]{verd10}
{Verdini}, A., M.~{Velli}, W.~H. {Matthaeus}, S.~{Oughton}, and P.~{Dmitruk},
\newblock {A Turbulence-Driven Model for Heating and Acceleration of the Fast
  Wind in Coronal Holes},
\newblock {\em Astrophys. J. Lett.}, {\em 708}, L116--L120, January 2010.

\end{thebibliography}

\end{document}